\pdfoutput=1
\documentclass[10pt]{article}

\usepackage{amsmath, fullpage}
\usepackage[left=2cm, right=2cm]{geometry}
\usepackage{titlesec}  

\usepackage{amssymb, amsthm}
\usepackage{subcaption}
\usepackage{graphicx}
\usepackage{authblk}

\usepackage[ruled]{algorithm2e}  

\usepackage[hidelinks]{hyperref}
\usepackage[numbers]{natbib}  

\usepackage{tikz}
\usepackage{cleveref}

\newtheorem{theorem}{Theorem}

\begin{document}
\font\myfont=cmr12 at 15pt
\title{\myfont Market Making of Options via Reinforcement Learning}
\author[1]{\small Zhou Fang}
\author[2]{\small Haiqing Xu}
\affil[1]{Department of Mathematics, The University of Texas at Austin}
\affil[2]{Department of Economics, The University of Texas at Austin}
\maketitle

\maketitle

\begin{abstract}
  Market making of options with different maturities and strikes is a challenging problem due to its highly dimensional nature. In this paper, we propose a novel approach that combines a stochastic policy and reinforcement learning-inspired techniques to determine the optimal policy for posting bid-ask spreads for an options market maker who trades options with different maturities and strikes.
\end{abstract}

\section{Introduction}
The option market maker plays a crucial role in the financial market, serving as a dealer that buys and sells options. Their role is to provide liquidity to the market by offering prices for options and by taking positions to manage the risk associated with their trades. However, the task of setting optimal prices for options with different strikes and maturities is highly non-trivial. In this paper, we address the challenging problem of market-making for options with multiple strikes and maturities.

The studies of market making start from \cite{grossman1988liquidity}
and \cite{ho1981optimal} in the 1980s. The idea in \cite{ho1981optimal} was revived in \cite{avellaneda2008high}, which inspired a large number of subsequent literature on market making. Notably, Cartea and Jaimungal \cite{cartea2014buy} and \cite{cartea2017algorithmic} have made influential contributions in this area. Other notable works include Baldacci et al. \cite{baldacci2021algorithmic}, Bergault and Breton \cite{bergault2021closed}, and Stoikov and Williams \cite{stoikov2009option}.

Reinforcement learning has been applied to market making in several works. For instance, Spooner and Lamba \cite{spooner2020robust}, Sadighian and Jaimungal \cite{sadighian2020extending}, Beysolow et al. \cite{beysolow2019market}, and Ganesh and Borkar \cite{ganesh2019reinforcement} have explored the use of reinforcement learning techniques in this context. However, these works often focus on engineering-oriented approaches and make simplifying assumptions in their models.

The use of stochastic policy is inspired by the reinforcement learning literature, and its first application in financial mathematics literature is found in Wang and Zhou \cite{wang2020continuous} for portfolio management problems. Stochastic policies offer improved robustness and strike a balance between exploration and exploitation.

Influential works such as those by Hu et al. (2013) \cite{hu2013kullback} and Jiang et al. (2016) \cite{jiang2016data} have examined uncertainty sets derived from Kullback-Leibler divergence distances and established a link with cross-entropy regularization. Drawing inspiration from these studies, we have adapted our original strategy to incorporate a stochastic policy enhanced by cross-entropy regularization. 

As a side note, there has been a recent surge in popularity for distributional robust optimization constrained by Wasserstein distance. Noteworthy contributions in this area include works by Esfahani et al. (2015) \cite{esfahani2015data}, Gao et al. (2022) \cite{gao2022wasserstein}, Abdullah et al. (2019) \cite{abdullah2019wasserstein}, and Hou et al. (2020) \cite{hou2020robust}. Although robust optimization and controls based on the Wasserstein distance hold great potential for future research, they are beyond the scope of this paper and will not be further explored here.

In this paper, we propose a reinforcement learning framework for market making in call options. We assume market order arrivals follow a Poisson process, with intensity inversely related to the bid-ask spreads. 

This paper is organized as follows. Section 2 introduces our model settings, while Section 3 presents the dynamic programming questions faced by the market maker, and the resulting Hamilton-Jacobi-Bellman equation. In Section 4, we prove a policy improvement theorem and the convergence result of the resulting policy-iteration algorithm. In Section 5, we state the policy-iteration algorithm in exact form. In Section 6, we conduct numerical experiments, and exhibit the final policy under different inventory levels and trading times for single and two options cases.

\section{Model Settings}


In this section, we present a market-making model for multiple related option assets, and its associated notation,  Throughout our discussion, we consider a probability space $(\Omega, \mathcal F, \mathbb P)$ accompanied by a filtration $(\mathcal F_t)_{t\in\mathbb N}$ satisfying the standard conditions.  Following \cite{baldacci2021algorithmic}, we  consider $n\geq 1$  call options written on a single underlying asset $S_t$ for $t\in [0,T]$. We assume that this asset follows  a geometric Brownian motion with non-stochastic coefficients over the interval $[0, T]$. Specifically, the dynamics of 
 $S_t$ are described by
\begin{align}
       \frac{dS_t}{S_t} = \sigma_t dW^S_t.
\end{align} where  $\sigma_t\in\mathbb R_+$ is a deterministic process, and $W^S_t$ is a standard Brownian motion process.  Note that we assume the volatility process $\sigma_t$ of $S_t$
to be non-stochastic. While this is a strong assumption made for the sake of simplicity in exposition, it is noteworthy that incorporating volatility's uncertainty would  influence the risk associated with its derivative assets. At the expense of more complicated notation, one could extend our analysis by including stochastic volatility to encapsulate the associated risk more comprehensively.

For  each option $i$, let $K_i \in \mathbb R_+$ and $T_i \in \mathbb R_+$ denote its  strike price and maturity, respectively. Throughout this discussion, we consider the market maker's decisions within the time period $[0,T]$, where $T< \min \{T_1,\cdots,T_n\}$. Furthermore, we assume the equilibrium mid-price of option $i$,  denoted by $P_{it}\in\mathbb R_+$, is expressed as follows:
\[
P_{it}=p_{it}(S_t, K_i, T_i)+\xi_{it}, \ \text{for} \ t\in[0,T].
\]Here, function $p_{it}$  represents the expected price of option $i$ given the current asset price $S_t$, i.e.,  $p_{it}\equiv \mathbb E(P_{it}|S_t)$, which is  also influenced by  a vector of deterministic  components, including the option strike price $K_i$, time to expiration $T_i$, and certain fundamental aspects capturing the dynamic process of $S_t$ (e.g. $\sigma_t$). Aligned with the Black-Scholes formula, the function $p_{it}$ can be further detailed using the  ``greeks''; see e.g. \cite{stoikov2009option}.   Moreover, the error term $\xi_{it}$ captures the influence attributable to supply and demand shocks in the option market, unrelated to the current stock price $S_t$. These shocks encompass market frictions like asymmetric information on the  asset's price process and market power, as discussed in e.g. \cite{kyle1985continuous}, among many others.

To measure the aggregate risk of the portfolio, we consider the correlations among the price processes of all $n$ options. Let $\boldsymbol \xi_t=(\xi_{1t},\cdots,\xi_{nt})^T \in\mathbb R^n$ represent the vector of shocks to the portfolio, $\boldsymbol p_t=(p_{1t},\cdots, p_{nt})^T \in\mathbb R^n$ denote the vector of expected prices, and $\boldsymbol P_t=(P_{1t},\cdots, P_{nt})^T \in\mathbb R^n$ the vector of mid-prices of the options, respectively,  at time $t$.  Specifically, we assume the dynamics of $\boldsymbol\xi_t$ are given by 
 \[
 d\boldsymbol \xi_{t}= \boldsymbol \Sigma_{t} d \boldsymbol W_{t},
 \]where $\boldsymbol \Sigma_{t}\in\mathbb R^{n\times n}$ is a deterministic process of  symmetric positive semidefinite matrices, and $\boldsymbol W_{t}\in\mathbb R^n$ is a standard $n$-dimensional vector of Brownian motions. The dynamic  process of $\boldsymbol \Sigma_t$ is critical as it describes the instantaneous correlations among the options' price shocks at any time $t$. Note that $\boldsymbol \Sigma_t$ is deterministic, implying that it does not depend on  the stochastic process $S_t$.  Therefore, by the Ito formula,  the dynamics of   $\boldsymbol P_{t}$ can be written as:
 \[
 d\boldsymbol P_{t}=\frac{\partial \boldsymbol p_{t}}{\partial t} dt+ \frac{\partial \boldsymbol p_{t}}{\partial S_t} dS_t+  \frac{\sigma_t^2 S_t^2 }{2} \frac{\partial^2 \boldsymbol p_{t}}{\partial S_t^2}  dt +\boldsymbol \Sigma_{t}d\boldsymbol W_t.
 \]
Again, here we assume $\boldsymbol \Sigma_t$ as a deterministic volatility process for the sake of simplification in expression.   In option theory, various risks are quantified using different "Greeks," which measure sensitivity to a variety of factors. In the dynamics of $\boldsymbol P_t$, $\frac{\partial \boldsymbol p_{t}}{\partial t}$ encapsulates the Theta risk, indicating the rate at which an option's price decreases as it nears expiration. $\frac{\partial \boldsymbol p_t}{\partial S_t}$ captures the Delta risk, reflecting the sensitivity of an option's price to changes in the underlying asset's price. Finally, $\frac{\partial^2 \boldsymbol p_{t}}{\partial S_t^2}$ represents the Vega risk, quantifying the impact of changes in volatility on the option’s price.
 
In the above framework, it's important to note that the correlation among the stochastic processes of the $n$ option prices can be decomposed into two distinct parts. The first element of correlation arises from the options' dependency on the same  asset  price $S_t$. The second arises from intrinsic market interdependencies, which may be  attributed to  demand, supply, or both sides, reflecting the degree of differentiation in the options' strike prices and expiration dates.

At each time  $t \in [0, T]$,  the market-maker sets the bid and ask prices for  option $i$ as  $P_{it} - \epsilon_{it}^-$ and $P_{it} + \epsilon_{it}^+$, respectively. Here, $\epsilon_{it}^-$ and $\epsilon_{it}^+$ denote the bid and ask spreads, both of which are non-negative.  Let $N_{it}^{+}\in\mathbb Z_+$ and $N_{it}^{-}\in\mathbb Z_+$ represent the counting processes for buy and sell market orders for call option $i$, respectively, and $Q_{it}$ as the inventory of call option $i$ at time $t$. Intuitively,  the dynamics of $Q_{it}$ is given by 
\begin{align}
dQ_{it} = dN_{it}^+ - dN_{it}^-.
\end{align} Following the literature on market making (see e.g. ),  we assume $N_{it}^{+}$ and $N_{it}^{-}$  follow Poisson processes with the  arrival intensities  $\lambda_{it}^+$ and $\lambda_{it}^-$, respectively, and given by
\[
\lambda_{it}^+\equiv \lambda_{it}^+(\epsilon^+_{it})=A_i^+\exp(-k_i^+\epsilon^+_{it});\ \ \ \
\lambda_{it}^-\equiv \lambda_{it}^-(\epsilon^-_{it})=A_i^-\exp(-k_i^-\epsilon^-_{it}),
\]where $A^+_i\in \mathbb R_+$ and $A^-_i\in \mathbb R_+$ is the constant frequency of market buys and sells, respectively;  $k_i^+\in\mathbb R_+$ and $k_i^-\in\mathbb R^+$ are two parameters that capture the idea that the demand of market orders depends on the bid and ask spreads, respectively. The intuition is that if the market-maker sets her quotes further from the mid-price, there is a reduced likelihood of the market executing against her buy or sell limit orders. For notation simplicity, we denote $\boldsymbol Q_t=(Q_{1t},\cdots, Q_{nt})^T\in\mathbb R^n$, $\boldsymbol N^+_t=(N^+_{1t},\cdots, N^+_{nt})^T \in\mathbb R^n_+$, $\boldsymbol N^-_t=(N^-_{1t},\cdots, N^-_{nt})'\in\mathbb R^n_+$,   $\boldsymbol \epsilon_t^-=(\epsilon_{1t}^-,\cdots,\epsilon_{nt}^-)^T\in\mathbb R^n_+$ and $\boldsymbol \epsilon_t^+=(\epsilon_{1t}^+,\cdots,\epsilon_{nt}^+)^T \in\mathbb R^n_+$. 

\[
N^+_{it}= m(\epsilon_{it}^+)
\]
Assumption: if
\[
\epsilon_{it}^+=0
\]then
\[
N^+_{it}=[\frac{\beta_1}{\epsilon_{it}^+}]
\]
%

To manage the risks inherent in options positions,  the market-maker is required to buy or sell  the underlying asset $S_t$ to offset or neutralize its impact  on the prices of options. In our analysis, we assume that the underlying asset market   is sufficiently liquid (i.e., it represents a complete market) to ensure a perfect $\Delta$-hedging.  Let $\Delta_t$ denote the position in shares that the market-maker holds at time $t$. Following e.g. \cite{stoikov2009option}, we have
\[
\Delta_t = \boldsymbol Q_t' \frac{\partial \boldsymbol p_{t}}{\partial S_t},
\]which  completely eliminates the risks associated with the price movements of $S_t$. This $\Delta$-hedging strategy  is a crucial component of the market-maker's decision, which is a maintained assumption  throughout our discussion. 

Let $X_t$ denote the Mark-to-Market value of a market-maker's portfolio, which includes cash, shares, and options. The dynamics of $X_t$ can be expressed as:
\[
    dX_t = dC_t - d( S_t\Delta_t ) + d(\boldsymbol P_{t}^T \boldsymbol Q_{t}), 
\]where $C_t$ represents the market-maker's cash process, described as
\[
 dC_t = (\boldsymbol \epsilon_{t}^+)^T d \boldsymbol N_{t}^{+} + (\boldsymbol \epsilon_{t}^-)^T d \boldsymbol N_{t}^- - \boldsymbol P_{t}^T d\boldsymbol Q_{t}   + S_t d\Delta_t +  d\langle \Delta, S \rangle_t.
\]From this, we derive the dynamics of $X_t$ as follows:
\begin{align}
    d X_t &= (\boldsymbol \epsilon_{t}^+)^T d\boldsymbol N_{t}^{+} + (\boldsymbol \epsilon_{t}^-)^T d\boldsymbol N_{t}^-  + \boldsymbol Q_t^T \big(\frac{\partial \boldsymbol p_{t}}{\partial t}  dt    + \frac{\partial \boldsymbol p_{t}}{\partial S_t} dS_t+  \frac{\sigma^2 S_t^2 }{2} \frac{\partial^2 \boldsymbol p_{t}}{\partial S_t^2}  dt + \boldsymbol \Sigma_{t}d \boldsymbol W_t \big)-\Delta_t dS_t\\
    &=(\boldsymbol \epsilon_{t}^+)^T d \boldsymbol N_{t}^{+} + (\boldsymbol \epsilon_{t}^-)^T d\boldsymbol N_{t}^-+ \boldsymbol Q_t^T \big(\frac{\partial \boldsymbol p_{t}}{\partial t}  dt    +   \frac{\sigma^2 S_t^2 }{2} \frac{\partial^2 \boldsymbol p_{t}}{\partial S_t^2}  dt \big)+ \boldsymbol Q^T_t \boldsymbol \Sigma_{t}d \boldsymbol W_t.
\end{align}In this equation, the term $(\boldsymbol \epsilon_{t}^+)^T d \boldsymbol N_{t}^{+} + (\boldsymbol \epsilon_{t}^-)^T d\boldsymbol N_{t}^-$ reflects direct returns from options transactions. Meanwhile, $ \boldsymbol Q_t^T \big(\frac{\partial \boldsymbol p_{t}}{\partial t}  dt    +   \frac{\sigma^2 S_t^2 }{2} \frac{\partial^2 \boldsymbol p_{t}}{\partial S_t^2}  dt \big)+ \boldsymbol Q^T_t \boldsymbol \Sigma_{t}d \boldsymbol W_t$ accounts for the inventory value of options and the underlying asset, divided into two segments:  $\boldsymbol Q_t^T \big(\frac{\partial \boldsymbol p_{t}}{\partial t}  dt    +   \frac{\sigma^2 S_t^2 }{2} \frac{\partial^2 \boldsymbol p_{t}}{\partial S_t^2}  dt \big)$ is a deterministic process capturing profit/loss from holding this option asset portfolio, and   $\boldsymbol Q^T_t \boldsymbol \Sigma_{t}d \boldsymbol W_t$ signifies the essential risk inherent in holding these assets, i.e., risks that cannot be offset by trading the underlying assets. It should also be noted that the position $\Delta_t$ of the underlying asset does not directly affect $X_t$'s dynamics, but it plays a crucial role in offsetting the risks of option positions due to uncertainty in the asset price $S_t$.

\section{Market-maker's Dynamic Programming (DP) Problem}

The objective of the market-maker is to maximize her expected payoff at the terminal time $T$. This involves determining the optimal bid/ask spread by solving a stochastic optimal control problem. In the existing market-making literature, it is commonly assumed that the market-maker's payoff is derived solely from the Mark-to-Market value of her entire portfolio at time $T$, rather than from its individual components (i.e. cash, shares and options). Moreover, it should also be noted that the residual positions in options also lead to costs associated with liquidation. In this context, we specifically consider the following dynamic programming problem
\[
\underset{(\epsilon^-_t,\epsilon^+_t): 0\leq t\leq T}{\max}\mathbb E [X_T-\psi(Q_T, \Delta_T, P_T, S_T)]
\]where $\psi(\cdot)$ represents the expected liquidation cost due to the inventory of the option profile. To simplify our analysis, we assume:
\[
\psi(Q_T, \Delta_T, P_T, S_T)=\psi_0\times \sum_{i=1}^n(P_{iT}Q_{iT})^2.
\]Here, $\psi_0>0$ is a positive coefficient that quantifies the impact of inventory on the market-maker's final value. Notably, the magnitude of 
$\psi_0$ is influenced by how complete the options market is, reflecting the way market characteristics affect associated liquidation costs. Because of the complete market assumption for the underlying asset, we assume the position $\Delta_T$  incurs no liquidation cost.

\subsection{Value function and policy function}

To solve the dynamic programming problem, we follow classical techniques of the stochastic optimal control method. First, we  introduce the value function and policy function.  The value function of the market maker can be defined as follows:  For $t\in[0,T]$, 
\begin{align}   
&V_t(\boldsymbol Q_t, \boldsymbol P_t, S_t) =\underset{(\boldsymbol \epsilon^-_s,\boldsymbol \epsilon^+_s): t\leq s\leq T}{\max} \mathbb E \left[\int_t^T dX_s-\psi_0 \times \sum_{i = 1}^{n} (P_{iT} Q_{iT})^2 \right] 
\end{align}
subject to
 \begin{eqnarray}
 \label{eq_6}
d\boldsymbol P_{t}&=&\frac{\partial \boldsymbol p_{t}}{\partial t} dt+ \frac{\partial \boldsymbol p_{t}}{\partial S_t} dS_t+  \frac{\sigma^2 S_t^2 }{2}\times \frac{\partial^2 \boldsymbol p_{t}}{\partial S_t^2}  dt + \boldsymbol V_{t}d \boldsymbol W_t \\
\label{eq_7}
d\boldsymbol Q_{t} &=& d \boldsymbol N^+_t-d \boldsymbol N^-_t;\\
\label{eq_8}
 d X_t  & =&(\boldsymbol \epsilon_{t}^+)^{T} d\boldsymbol N_t^+ + (\boldsymbol \epsilon_{t}^-)^{T} d\boldsymbol N^-_t+ \boldsymbol Q_t^{T} \big(\frac{\partial \boldsymbol p_{t}}{\partial t}  dt    +   \frac{\sigma^2_t S_t^2 }{2} \frac{\partial^2 \boldsymbol p_{t}}{\partial S_t^2}  dt \big)+ \boldsymbol Q^{T}_t \boldsymbol V_{t}d \boldsymbol W_t.
\end{eqnarray}
Suppose the transversality condition holds, then the value function satisfies the Hamilton-Jacobi-Bellman (HJB) equation (see the derivation in the appendix~\ref{HJB deterministic}):
\begin{multline*}
V_t(\boldsymbol Q_t, \boldsymbol P_t, S_t)  \\
=\lim_{\Delta t\to 0} \ \ \underset{(\boldsymbol \epsilon_s^+,\boldsymbol \epsilon_s^-): t\leq s\leq t+\Delta t}{\max} \left\{\mathbb E \Big[ \int_t^{t+\Delta t} dX_s \ \Big | \ \boldsymbol Q_{t}, \boldsymbol P_{t}, S_{t} \Big] +  \mathbb E \left[  V_{t+\Delta t}(\boldsymbol Q_{t+\Delta t}, \boldsymbol P_{t+\Delta t} , S_{t+\Delta t}) \hspace{0.1cm}\Big |\hspace{.1cm}\boldsymbol Q_{t}, \boldsymbol P_{t}, S_{t}  \right] \right\}.
\end{multline*}
Plugging equations (\ref{eq_6}) to (\ref{eq_8}) into the RHS,  the above HJB  can be written as 
\begin{multline*}
0=\underset{\boldsymbol \epsilon_t^+,\boldsymbol \epsilon_t^-\in\mathbb R^n_+}{\max} \ \Big\{\sum_{i=1}^n\epsilon_{it}^+ \times  \lambda_{it}^+(\epsilon^+_{it}) + \sum_{i=1}^n \epsilon_{it}^-  \times \lambda^-_{it}(\epsilon^-_{it})+ \sum_{i = 1}^{n} \frac{\partial V_{t}(\boldsymbol{Q}_t , \boldsymbol{P}_t, S_t)}{\partial Q_{it}}\left[\lambda_{it}^+(\epsilon_{it}^+)-\lambda_{it}^-(\epsilon_{it}^-)\right] \Big\} \\
+ \ \boldsymbol Q_t^{T} \big(\frac{\partial \boldsymbol p_{t}}{\partial t}    +   \frac{\sigma_t^2 S_t^2 }{2} \frac{\partial^2 \boldsymbol p_{t}}{\partial S_t^2}  \big) 
+ \frac{\partial V_t(\boldsymbol{Q}_t, \boldsymbol{P}_t, S_t)}{\partial t} 
+ \frac{\sigma^2_t S_t^2}{2} \frac{\partial^2 V_t(\boldsymbol{Q}_t, \boldsymbol{P}_t, S_t)}{\partial S^2_t}   + (\nabla_{\boldsymbol{P}} V_t) \Delta \boldsymbol{P} + \frac{1}{2}(\Delta \boldsymbol{P})^{T} \boldsymbol{H}_{\boldsymbol{P}} \Delta \boldsymbol{P}. 
\end{multline*}
In the above expression, $\nabla V_t$ and $\boldsymbol{H_P}$ are defined as follows:
\begin{align*}
    \nabla_{\boldsymbol{P}} V_t = \Big( \frac{V_t(\boldsymbol Q_t, \boldsymbol{ P_t, S_t})}{\partial P_{t}}, \cdots, \frac{V_t(\boldsymbol Q_t, \boldsymbol{ P_t, S_t})}{\partial P_{t}} \Big);\\
\mathbf{H_P} = 
\begin{bmatrix}
    \frac{\partial^2 V_t(\boldsymbol Q_t, \boldsymbol{ P_t, S_t})}{\partial P_1^2} & \cdots & \frac{\partial^2 V_t(\boldsymbol Q_t, \boldsymbol{ P_t, S_t})}{\partial P_1 \partial P_n} \\
    \vdots & \cdots & \dots
    \\
    \frac{\partial^2 V_t(\boldsymbol Q_t, \boldsymbol{ P_t, S_t})}{\partial P_n \partial P_1} &\cdots & \frac{\partial^2 V_t(\boldsymbol Q_t, \boldsymbol{ P_t, S_t})}{\partial P_n^2}
\end{bmatrix}.
\end{align*}
Note that we apply the Ito formula for the  derivation of the above HJB equation, in particular, we treat the  inventory $Q_t$ as a continuous random variables for the simplicity of notation. The exact HJB equation under deterministic policy is 
\begin{align}
    &\max_{\boldsymbol \epsilon_t^+, \boldsymbol \epsilon_t^- \in \mathbb R^{n}_+} \Bigg\{  \sum_{i = 1}^{n} \lambda_{it} \big(  \epsilon_{it}^+ + \Delta_i^+ V_t(\boldsymbol{Q}_t, \boldsymbol{P}_t, S_t)  \big) + \sum_{i = 1}^{n} \lambda_{it} \big( \epsilon_{it}^- + \Delta_i^- V_t(\boldsymbol{Q}_t, \boldsymbol{P}_t, S_t) \big) \Bigg\} \nonumber \\
    &+ \boldsymbol{Q}_t^{T} \big( 
    \frac{\partial \boldsymbol{p_t}}{\partial t} + \frac{\partial^2 \boldsymbol p_t}{\partial S_t^2}\big) + \partial_t V_t( \boldsymbol{Q}_t, \boldsymbol{P}_t, S_t) + \frac{\sigma^2 S_t^2}{2}\partial_{SS} V_t( \boldsymbol{Q_t}, \boldsymbol{P}_t, S_t) \nonumber \\
    &+ (\nabla_{\boldsymbol{P}} V_t)^{T} \big( \frac{\partial \boldsymbol{p}_t}{\partial t}  + \frac{\sigma^2S_t^2}{2} \frac{\partial^2 \boldsymbol{p}_t}{\partial S_t^2} \big) + \Big[\frac{\sigma^2 S_t^2}{2} (\frac{\partial \boldsymbol{p}_t}{\partial S_t})^T \boldsymbol{H}_{\boldsymbol{P}}   (\frac{\partial \boldsymbol{p}_t}{\partial S_t}) + \frac{1}{2} \textbf{Tr} (\boldsymbol{V}_t^{T} \boldsymbol{H}_{\boldsymbol{P}} \boldsymbol{V}_t) \Big] = 0
\end{align}

where, 
\begin{align}
    \Delta_i^+ V_t(\boldsymbol{Q}_t, \boldsymbol {P}_t, S_t) &= V_t(\boldsymbol{Q}_t + \boldsymbol{\Delta}_i, \boldsymbol{P}_t, S_t) - V_t(\boldsymbol{Q}_t, \boldsymbol{P}_t, S_t), \\
    \Delta_i^- V_t(\boldsymbol{Q}_t, \boldsymbol{P}_t, S_t) &= V_t(\boldsymbol{Q}_t - \boldsymbol{\Delta}_i, \boldsymbol{P}_t, S_t) - V_t(\boldsymbol{Q}_t, \boldsymbol{P}_t, S_t).
\end{align}

\subsection{Uncertainty on Market Parameters and Exploratory Optimization}
In practice, the market parameters in the above Market Maker’s DP problem are fundamentally unknown or misspecified, such as the dynamics of $S_t$, $\boldsymbol P_{t}$, and the policy-contingent Poisson processes $(\boldsymbol N^-_t, \boldsymbol N^+_t)$. In principle, it is feasible to estimate these market parameters from training samples and then plug these estimates into the HJB equation to solve a fixed point for the value function, which subsequently leads to an approximate optimal policy solution.  However, these solutions often inherit any estimation errors and model misspecification biases from the initial step. This issue is known as the optimizer's curse in decision theory. Moreover, the task of solving the fixed point in the high-dimensional space of value functions is known to be extremely sensitive to these parameters and their specific forms of parametrization, as discussed in e.g. \cite{michaud1989markowitz}.

Therefore, we propose a robust optimal solution by applying a recently developed reinforcement learning approach, augmented with cross-entropy regularization. This approach does not require accurate information on market parameters, but accommodates their uncertainties  and identifies the optimal policy through a process of exploration and trial-and-error.  Employing the HJB framework, it continuously refines policy in line with the market maker's objectives, bypassing the need to solve the fixed point of the HJB equation for the value function. While highly promising, it's important to recognize potential challenges in applying RL in such complex environments, particularly in effectively balancing the exploration-exploitation tradeoff to enable the reinforcement learning model to accurately approximate the optimal solution. 

Following the literature \cite{wang2020continuous}, our reinforcement learning algorithm adopts  mixed strategies (i.e. stochastic policies) to facilitate exploration and approximate the optimal solution.  Unlike pure strategy policies, which always select a single buy and sell spread in a given state, stochastic policies randomly distribute probabilities across a range of spreads.  The randomization enables the selection of less obvious pricing policies, enhancing the chance of finding optimal or near-optimal solutions and preventing from getting stuck in suboptimal buy and sell spreads.  In financial markets, an additional benefit of employing a stochastic policy is its strategyproofness, i.e., the uncertainty introduced by the mixed strategy makes market timing much more difficult for  market participants.

Let $\pi_t$ be a probability density function supported on $\mathbb R_+^2$, which represents the randomized bid/ask spreads  as a mixed strategy at time $t$.  To integrate the  ``exploratory'' mechanism into the Market maker's optimization problem, we reformulated  the value function  with the cross-entropy regularization as follows: 
\begin{align}
    &V_t(\boldsymbol Q_t, \boldsymbol P_t, S_t) \nonumber \\
    =& \underset{(\epsilon^-_s,\epsilon^+_s): t\leq s\leq T}{\max}\  \mathbb E \Bigg[\int_s^T d \widetilde{X}_s-\psi_0 \sum_i^n (P_{iT} Q_{iT})^2 \nonumber \\
    -&  \ \gamma \int_s^T \int_{\boldsymbol{\epsilon}} \boldsymbol{\pi}(\boldsymbol{\epsilon}_t^{\pm} | \boldsymbol{Q}_u, \boldsymbol{P}_u, S_u) \log  \boldsymbol{\pi}(\boldsymbol{\epsilon}_t^{\pm} | \boldsymbol{Q}_u, \boldsymbol{P}_u, S_u) d\boldsymbol{\epsilon}_u du\Bigg]  \nonumber 
\end{align}
where
\begin{align}
    d \widetilde{X}_t = \int_{\boldsymbol{\epsilon}_t} \boldsymbol{\pi}(\epsilon_t) \Big[ (\boldsymbol \epsilon_{t}^+)^{T} d\boldsymbol N_t^+ + (\boldsymbol \epsilon_{t}^-)^{T} d\boldsymbol N^-_t\Big] d\boldsymbol{\epsilon}_t  + \boldsymbol Q_t^{T} \big(\frac{\partial \boldsymbol p_{t}}{\partial t}  dt    +   \frac{\sigma_t^2 S_t^2 }{2} \frac{\partial^2 \boldsymbol p_{t}}{\partial S_t^2}  dt \big)+ \boldsymbol Q^{T}_t \boldsymbol V_{t}d \boldsymbol W_t. \nonumber 
\end{align}
where $\boldsymbol{\epsilon}_s \equiv (\boldsymbol{\epsilon}^-_s,\boldsymbol{\epsilon} ^+_s )\in\mathbb R^2_+$, $\Pi$ is the class of admissible density-function-valued  (and supported on $\mathbb R^2_+$) stochastic processes, and $\gamma\in\mathbb R_+$ is a positive tuning parameter.  It follows that the corresponding HJB equation takes the following expression:  
\begin{align}
0=&\underset{\pi_t\in\Pi_t}{\max} \Bigg\{ \int \Big\{{\boldsymbol\epsilon'}_{t}^+  \boldsymbol \lambda_{t}^+(\boldsymbol\epsilon^+_{t}) + {\boldsymbol\epsilon'}_{t}^-  \boldsymbol\lambda^-_{t}(\boldsymbol\epsilon^-_{t})+  \frac{\partial V_{t}(\boldsymbol{Q}_t , \boldsymbol{P}_t, S_t)}{\partial \boldsymbol Q'_{t}}\left[\boldsymbol\lambda_{t}^+(\boldsymbol\epsilon_{t}^+)-\boldsymbol\lambda_{t}^-(\boldsymbol\epsilon_{t}^-)\right] \Big\} \pi_t(\boldsymbol{\epsilon}_t| \boldsymbol{Q}_t, \boldsymbol{P}_t, S_t) d\boldsymbol{\epsilon}_t   \nonumber \\
-& \ \gamma   \int \boldsymbol{\pi}_t(\boldsymbol{\epsilon}_t | \boldsymbol{Q}_t, \boldsymbol{P}_t, S_t) \log  \boldsymbol{\pi}_t(\boldsymbol{\epsilon}_t | \boldsymbol{Q}_t, \boldsymbol{P}_t, S_t) d\boldsymbol{\epsilon}_t \Bigg\} + \boldsymbol Q_t^{T} \big(\frac{\partial \boldsymbol p_{t}}{\partial t}  +   \frac{\sigma_t^2 S_t^2 }{2} \frac{\partial^2 \boldsymbol p_{t}}{\partial S_t^2}  \big) \nonumber  \\
+& \  \frac{\partial V_t(\boldsymbol{Q}_t, \boldsymbol{P}_t, S_t)}{\partial t}  + \frac{\sigma_t^2S_t^2}{2} \frac{\partial^2 V_t(\boldsymbol{Q}_t, \boldsymbol{P}_t, S_t)}{\partial S^2_t}   + (\nabla_{\boldsymbol{P}} V_t) \Delta \boldsymbol{P} + \frac{1}{2}(\Delta \boldsymbol{P})^{T} \boldsymbol{H}_{\boldsymbol{P}} \Delta \boldsymbol{P}. 
\end{align}

The above equation simplifies the notation as in the HJB equation under deterministic policy, the exploratory HJB equation is as follows. See \cite{wang2020continuous} for detailed introduction to exploratory HJB equation. 
\begin{align}
    &\max_{\boldsymbol{\pi}} \Bigg\{ \int_{\boldsymbol{\epsilon}_t} \boldsymbol{\pi}(\boldsymbol{\epsilon_t} | \boldsymbol{Q}_t, \boldsymbol{P}_t, S_t)  \sum_{i = 1}^{n} \bigg( \lambda_{it} \big[  \epsilon_{it}^+ + \Delta_i^+ V_t(\boldsymbol{Q}_t, \boldsymbol{P}_t, S_t) \big] + \lambda_{it} \big[ \epsilon_{it}^- + \Delta_i^- V_t(\boldsymbol{Q}_t, \boldsymbol{P}_t, S_t)  \big] \bigg) d\boldsymbol{\epsilon_t} \nonumber \\
    & - \gamma \int_{\boldsymbol{\epsilon}} \boldsymbol{\pi}(\boldsymbol{\epsilon}_t^{\pm} | \boldsymbol{Q}_t, \boldsymbol{P}_t, S_t) \log \boldsymbol{\pi}(\boldsymbol{\epsilon}_t^{\pm} | \boldsymbol{Q}_t, \boldsymbol{P}_t, S_t) d\boldsymbol{\epsilon}
    \Bigg\}  \nonumber \\
    &+ \boldsymbol{Q}_t^{T} \big( \frac{\partial \boldsymbol{p_t}}{\partial t} + \frac{\partial^2 \boldsymbol p_t}{\partial S_t^2}\big) + \partial_t V_t( \boldsymbol{Q}_t, \boldsymbol{P}_t, S_t) + \frac{\sigma^2 S_t^2}{2}\partial_{SS} V_t( \boldsymbol{Q_t}, \boldsymbol{P}_t, S_t) \nonumber \\
    &+ (\nabla_{\boldsymbol{P}} V_t)^{T} \big( \frac{\partial \boldsymbol{p}_t}{\partial t}  + \frac{\sigma_t^2}{2} \frac{\partial^2 \boldsymbol{p}_t}{\partial S_t^2} \big) + \Big[\frac{\sigma_t^2 S_t^2}{2} (\frac{\partial \boldsymbol{p}_t}{\partial S_t})^T \boldsymbol{H}_{\boldsymbol{P}}   (\frac{\partial \boldsymbol{p}_t}{\partial S_t}) + \frac{1}{2} \textbf{Tr} (\boldsymbol{V}_t^{T} \boldsymbol{H}_{\boldsymbol{P}} \boldsymbol{V}_t) \Big] = 0
\end{align}

\section{Optimal Policy}
Upon deriving the exploratory Hamilton-Jacobi-Bellman (HJB) equation, we identify the requisite condition that the optimal policy must fulfill, as outlined below. 
\begin{align}
     \boldsymbol{\pi}^* \sim \frac{\exp \Big \{ \frac{1}{\gamma} \sum_{i = 1}^{n} \lambda_{it}^+ \big(  \epsilon_{it}^+ + \Delta_i^+ V_t(\boldsymbol{Q}_t, \boldsymbol{P}_t, S_t) \big) + \sum_{i = 1}^{n} \lambda_{it}^- \big( \epsilon_{it}^- + \Delta_i^- V_t (\boldsymbol{Q}_t, \boldsymbol{P}_t, S_t)\big)    \Big\}}{\int_{\boldsymbol{\epsilon_t}}\exp \Big \{ \frac{1}{\gamma} \sum_{i = 1}^{n} \lambda_{it}^+ \big(  \epsilon_{it}^+ + \Delta_i^+ V_t(\boldsymbol{Q}_t, \boldsymbol{P}_t, S_t) \big) + \sum_{i = 1}^{n} \lambda_{it}^- \big( \epsilon_{it}^- + \Delta_i^- V_t(\boldsymbol{Q}_t, \boldsymbol{P}_t, S_t)\big)    \Big\} d\boldsymbol{\epsilon_t}}
\end{align}
The derivation process entails demonstrating that the optimal policy is an 'interior' point within the policy set. This involves applying calculus of variations techniques, details of which are provided in Appendices~\ref{concavity of maximization} and~\ref{Derive Optimal Policy}

While the optimal policy is subject to a specific condition involving the value function, this leads to an intractable partial differential equation of the value function. In this project, we introduce a reinforcement learning framework that allows us to begin with a random policy and approximate the optimal policy. Prior to detailing the reinforcement learning algorithm, we present the following two theorems that form the foundation of our reinforcement learning algorithm.

The first theorem is the Policy Improvement Theorem, which outlines a methodology for deriving an enhanced policy from an existing one. This theorem serves as the cornerstone for policy improvement within our reinforcement learning algorithm, offering a systematic approach to refine policies iteratively towards optimality.

\begin{theorem}[\textbf{policy improvement theorem}]
     Given arbitrary admissible policy $\boldsymbol{\pi}$, the value function under the policy is $V^{\boldsymbol{\pi}}$. There is a new policy derived from this policy, defined as 
     \begin{align}
         \widetilde{\boldsymbol{\pi}} \sim \frac{\exp \Big \{ \frac{1}{\gamma} \sum_{i = 1}^{n} \lambda_{it}^+ \big(  \epsilon_{it}^+ + \Delta_i^+ V_t^{\boldsymbol{\pi}}(\boldsymbol{Q}_t, \boldsymbol{P}_t, S_t) \big) + \sum_{i = 1}^{n} \lambda_{it}^- \big( \epsilon_{it}^- + \Delta_i^- V_t^{\boldsymbol{\pi}}(\boldsymbol{Q}_t, \boldsymbol{P}_t, S_t)\big)    \Big\}}{\int_{\boldsymbol{\epsilon_t}}\exp \Big \{ \frac{1}{\gamma} \sum_{i = 1}^{n} \lambda_{it}^+ \big(  \epsilon_{it}^+ + \Delta_i^+ V_t^{\boldsymbol{\pi}}(\boldsymbol{Q}_t, \boldsymbol{P}_t, S_t) \big) + \sum_{i = 1}^{n} \lambda_{it}^- \big( \epsilon_{it}^- + \Delta_i^- V_t^{\boldsymbol{\pi}}(\boldsymbol{Q}_t, \boldsymbol{P}_t, S_t)\big)    \Big\} d\boldsymbol{\epsilon_t}}
     \end{align}
     Then, there is the following inequality holds for all $(t, \boldsymbol{Q}_t, \boldsymbol{P}_t, S_t)$
     \begin{align}
         V^{\boldsymbol{\pi}}_t(\boldsymbol{Q}_t, \boldsymbol{P}_t, S_t) \leq V^{\boldsymbol{\widetilde{\pi}}}_t(\boldsymbol{Q}_t, \boldsymbol{P}_t, S_t) 
     \end{align}
\end{theorem}
The detailed proof is in appendix~\ref{Policy Improvement}. 

Merely relying on the Policy Improvement Theorem is insufficient because it does not guarantee that continual policy updates will converge to the optimal policy rather than settling on a sub-optimal policy, which also represents a fixed point. It's important to note that in traditional reinforcement learning literature, the presence of a discounting factor and the framing of problems within an infinite-time horizon are common. These elements introduce considerations about when and how convergence to the optimal policy is achieved, underscoring the need for additional theoretical support to ensure that policy iteration leads to the optimal solution.

\begin{theorem}[\textbf{convergence to optimal}]
    Given any policy $\boldsymbol{\pi}_0$, one can generate a sequences of improved policies, $\boldsymbol{\pi}_1, \boldsymbol{\pi}_2, ..., \boldsymbol{\pi}_n, ...$. Then the value functions converge, denoted as $V^{\infty}$. There is a sequence of inequalities, 
    \begin{align}
        V^{\boldsymbol{\pi}_1} \leq V^{\boldsymbol{\pi}_2} \leq \dots \leq V^{\boldsymbol{\pi}_n} \leq \dots \leq V^{\infty}
    \end{align}
    We claim that it converges to the optimal value function, thus, the policy also converges to one optimal policy.
    \begin{align}
        V^{\infty} = V
    \end{align}
\end{theorem}
The detailed proof is in appendix~\ref{Optimal Convergence}. Armed with the assurance of convergence, we are positioned to apply the Policy Improvement Theorem iteratively, enabling us to approximate the optimal policy as closely as desired. This iterative application ensures a methodical enhancement of policies, guiding us towards optimal decision-making strategies over time.

\section{Reinforcement Learning Algorithm}

In this section, we propose a reinforcement learning algorithm following the spirit of policy improvement theorem and define the continuous-time version of temporal-difference errors. 

Assume current policy is $\boldsymbol{\pi}$, then the following equation holds
\begin{align}
    V_t^{\boldsymbol{\pi}}(\boldsymbol{Q}_t, \boldsymbol{P}_t, S_t) &= \mathbb E \Big[ \int_t^s \int_{\boldsymbol{\epsilon}} \boldsymbol{\pi}(\boldsymbol{\epsilon}_u) \sum_{i = 1}^{n} \big( \epsilon_{iu}^+ dN_{iu}^+  + \epsilon_{iu}^- dN_{iu}^- \big) d\boldsymbol{\epsilon}_u du - \gamma \int_t^s \int_{\boldsymbol{\epsilon}_u} \boldsymbol{\pi}(\boldsymbol{\epsilon}_u) \log \boldsymbol{\pi}(\boldsymbol{\epsilon}_u) d\boldsymbol{\epsilon}_u du \nonumber \\ 
        &+ \int_t^s  \boldsymbol{Q}_u^{T} \big( \frac{\partial \boldsymbol{p}_u}{\partial t} + \frac{\sigma^2 S_t^2}{2} \frac{\partial^2 \boldsymbol p_u}{\partial S^2}\big) du + V^{\boldsymbol{\pi}}_s(\boldsymbol{Q}_s, \boldsymbol{P}_s, S_s) \ \Big | \ \boldsymbol{Q}_t, \boldsymbol{P}_t, S_t \Big]
\end{align}
which leads to
\begin{align}
    0&= \mathbb E \Big[ \frac{1}{s - t} \int_t^s \int_{\boldsymbol{\epsilon}} \boldsymbol{\pi}(\boldsymbol{\epsilon}_u) \sum_{i = 1}^{n} \big( \epsilon_{iu}^+ dN_{iu}^+  + \epsilon_{iu}^- dN_{iu}^- \big) d\boldsymbol{\epsilon}_u du - \frac{\gamma}{s - t} \int_t^s \int_{\boldsymbol{\epsilon}_u} \boldsymbol{\pi}(\boldsymbol{\epsilon}_u) \log \boldsymbol{\pi}(\boldsymbol{\epsilon}_u) d\boldsymbol{\epsilon}_u du \nonumber \\ 
        &+ \frac{1}{s - t} \int_t^s  \boldsymbol{Q}_u^{T} \big( \frac{\partial \boldsymbol{p}_u}{\partial t} + \frac{\sigma^2 S_t^2}{2} \frac{\partial^2 \boldsymbol p_u}{\partial S^2}\big) du + \frac{V^{\boldsymbol{\pi}}_s(\boldsymbol{Q}_s, \boldsymbol{P}_s, S_s) - V_t^{\boldsymbol{\pi}}(\boldsymbol{Q}_t, \boldsymbol{P}_t, S_t) }{s - t}  \ \Big | \ \boldsymbol{Q}_t, \boldsymbol{P}_t, S_t \Big] \nonumber 
\end{align}

The first step is to estimate the value function under policy $\boldsymbol{\pi}$. Assume there is a parametric (usually a neural network) for the value function, defined as $V^{\boldsymbol{\pi}}(\boldsymbol{Q}, \boldsymbol{P}, S ; \theta)$. We define the temporal difference in continuous time as follows

\begin{align}
    \delta_t^{\theta} &= \lim_{s \to t } \  \mathbb{E} \Big[\frac{V^{\boldsymbol{\pi}}_s(\boldsymbol{Q}_s, \boldsymbol{P}_s, S_s ; \theta) - V_t^{\boldsymbol{\pi}}(\boldsymbol{Q}_t, \boldsymbol{P}_t, S_t ; \theta) }{s - t}  \ \Big | \ \boldsymbol{Q}_t, \boldsymbol{P}_t, S_t
    \Big]  \nonumber \\
    &+ \int_{\boldsymbol{\epsilon}} \boldsymbol{\pi}(\boldsymbol{\epsilon}_t | \boldsymbol{Q}_t, \boldsymbol{P}_t, S_t) \sum_{i = 1}^{n} \big( \lambda_{it}^+ \epsilon_{it}^+   + \lambda_{it}^- \epsilon_{it}^- \big) d\boldsymbol{\epsilon}_t  +  \boldsymbol{Q}_u^{T} \big( \frac{\partial \boldsymbol{p}_u}{\partial t} + \frac{\sigma^2 S_t^2}{2} \frac{\partial^2 \boldsymbol p_u}{\partial S^2}\big) \nonumber \\ 
    & -  \gamma \int_{\boldsymbol \epsilon}  \boldsymbol \pi(\boldsymbol \epsilon_t | \boldsymbol{Q}_t, \boldsymbol{P}_t, S_t) \log  \boldsymbol \pi(\boldsymbol \epsilon_t | \boldsymbol{Q}_t, \boldsymbol{P}_t, S_t ) d\boldsymbol \epsilon_t
\end{align}
the loss function needs to be minimized is
\begin{align}
    \mathcal{L}(\theta) = \mathbb{E} \big[  \int_0^T |\delta_t^{\theta} | ^ 2 dt \big]
\end{align}

Of course, when conducting the numerical analysis, we need to discretize time and minimize the loss function accordingly. The following is a summary of the policy iteration algorithm 1.
\begin{algorithm}
    \caption{Policy Iteration with Neural Network Approximation}
    \label{alg:policy_iteration}  
    \SetAlgoNoLine
    \KwIn{Initialize hyperparameters, learning rate $\alpha$, and neural network parameters $\theta$.}
    \KwOut{Optimized policy $\boldsymbol{\pi}^*$.}

    \For{$l = 1$ to $L$}{
        \For{$d = 1$ to $D$}{
            Generate one sample path under policy $\boldsymbol{\pi}$\;
        } 
        Compute loss function $\mathcal{L}(\theta)$\;
        Update neural network parameters via gradient descent:
        \[
            \theta \leftarrow \theta - \alpha \nabla_{\theta} \mathcal{L}(\theta)
        \] 
        Update policy using softmax transformation:
        \[
            \boldsymbol{\pi} \sim 
            \frac{\exp \left( \frac{1}{\gamma} \sum\limits_{i = 1}^{n} 
            \lambda_{it}^+ \big( \epsilon_{it}^+ + \Delta_i^+ V_t^{\boldsymbol{\pi}}(\boldsymbol{Q}_t, \boldsymbol{P}_t, S_t ; \theta) \big) 
            + \sum\limits_{i = 1}^{n} \lambda_{it}^- \big( \epsilon_{it}^- + \Delta_i^- V_t^{\boldsymbol{\pi}}(\boldsymbol{Q}_t, \boldsymbol{P}_t, S_t ; \theta) \big)  
            \right)}
            {\int_{\boldsymbol{\epsilon_t}} \exp \left( \frac{1}{\gamma} \sum\limits_{i = 1}^{n} 
            \lambda_{it}^+ \big( \epsilon_{it}^+ + \Delta_i^+ V_t^{\boldsymbol{\pi}}(\boldsymbol{Q}_t, \boldsymbol{P}_t, S_t ; \theta) \big) 
            + \sum\limits_{i = 1}^{n} \lambda_{it}^- \big( \epsilon_{it}^- + \Delta_i^- V_t^{\boldsymbol{\pi}}(\boldsymbol{Q}_t, \boldsymbol{P}_t, S_t ; \theta) \big)  
            \right) d\boldsymbol{\epsilon_t}}
        \]
    }
\end{algorithm}

\section{Numerical Experiments}

We conduct a simple numerical experiment on a single-option case to examine whether the final policy aligns with standard option market-making intuition. Specifically, as inventory increases, the bid price should rise while the ask price should decrease, since the market maker seeks to offset inventory risk.

\begin{figure}[htbp]
    \centering
    \begin{subfigure}[b]{0.45\linewidth}
        \centering
        \includegraphics[width=\linewidth]{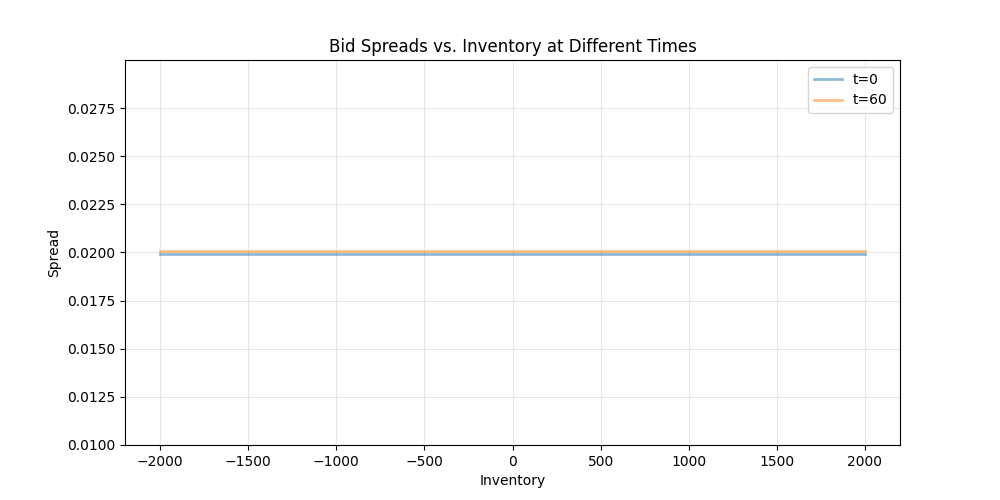}
        \caption{Bid spread under different inventory levels}
        \label{fig:bid-inv_no_penalty}
    \end{subfigure}
    \hfill
    \begin{subfigure}[b]{0.45\linewidth}
        \centering
        \includegraphics[width=\linewidth]{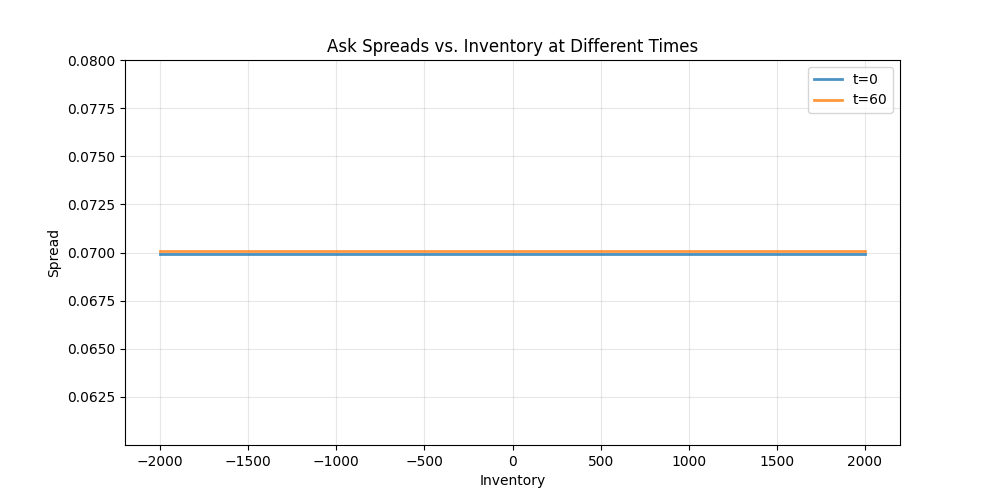}
        \caption{Ask spread under different inventory levels}
        \label{fig:ask-inv_no_penalty}
    \end{subfigure}
    \caption{Bid-ask spreads under different inventory levels with no terminal penalty}
    \label{fig:bid-ask-no_penalty}
\end{figure}

Figures \ref{fig:bid-inv_no_penalty} and \ref{fig:ask-inv_no_penalty} illustrate the bid and ask spreads for \( t = 0 \) and \( t = 60 \), given a fixed stock price and varying inventory levels. The left subfigure shows the bid spreads, while the right subfigure presents the ask spreads. The hyperparameters for these figures are set to \( \psi_0 = 0 \) (indicating no penalty for holding inventory at the end of the trading period) and \( \gamma = 0 \), meaning the policy is deterministic.

As observed, the policy plots for different trading periods (\( t = 0 \) and \( t = 60 \)) remain identical. This result aligns with expectations: since there is no penalty for holding inventory at the final trading period, the policy at each step solely focuses on maximizing wealth, including the value of the inventory. Consequently, inventory levels do not influence the bid-ask policy.

\begin{figure}[htbp]
    \centering
    \includegraphics[width=\linewidth]{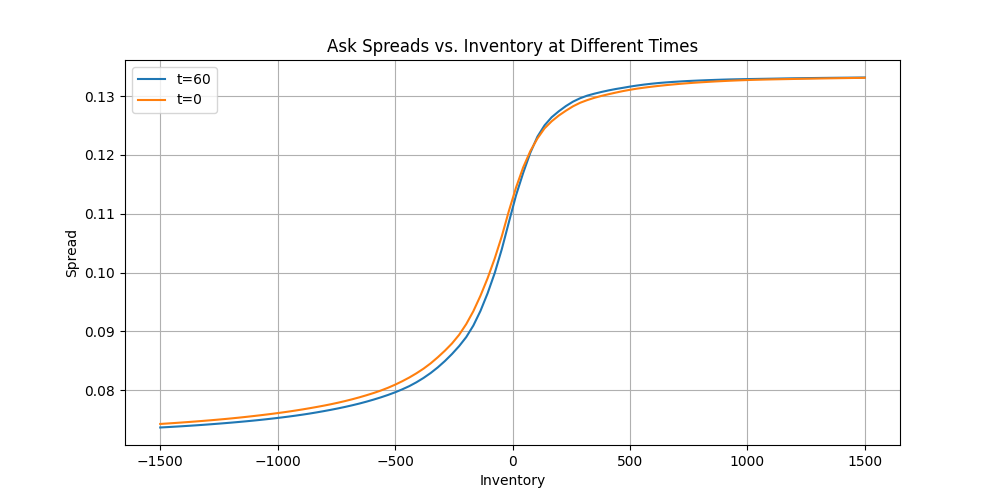}
    \caption{Bid spreads under different inventory levels for \( t = 0 \) and \( t = 60 \).}
    \label{bid_inv_multi_time}
\end{figure}

\begin{figure}[htbp]
    \centering
    \includegraphics[width=\linewidth]{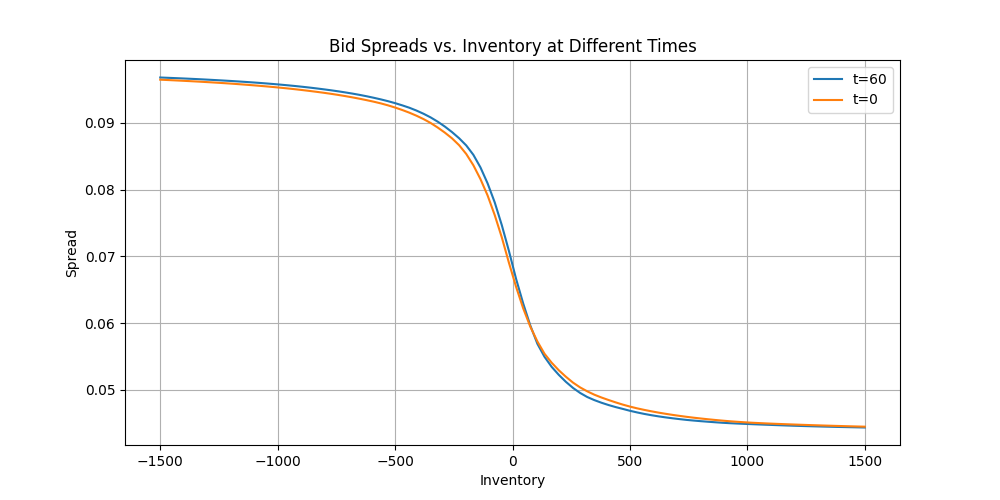}
    \caption{Ask spreads under different inventory levels for \( t = 0 \) and \( t = 60 \).}
    \label{ask_inv_multi_time}
\end{figure}

The previous case served as a sanity check to ensure that the model’s output aligns with our intuition. Now, we extend the analysis to the case where \( \psi_0 > 0 \), meaning there is a penalty for holding inventory at the end of the trading period. Figures \ref{bid_inv_multi_time} and \ref{ask_inv_multi_time} illustrate the bid and ask spreads, respectively, at two different time steps (\( t = 0 \) and \( t = 60 \)).  

A key observation is that as inventory levels change, the bid and ask spread curves exhibit significant variation. A general pattern across all trading periods is that an **increase in inventory leads to a widening of the ask spread and a narrowing of the bid spread**. This behavior is intuitive: when inventory increases, the market maker adjusts the bid-ask spreads strategically to encourage inventory reduction. Specifically, the increase in the ask spread implies a rise in the ask price, while the decrease in the bid spread suggests an increase in the bid price. These adjustments make it **more attractive to buy from the market and less attractive to sell**, thus facilitating inventory offsetting in line with optimal market-making behavior.

A closer look at the bid-ask curves for \( t = 0 \) and \( t = 60 \) reveals an important nuance: the magnitude of the spread adjustment is larger when the trading period is nearing its end. In other words, as inventory increases, the bid-ask spread curves for \( t = 60 \) exhibit steeper changes compared to those at \( t = 0 \). This finding aligns with intuition, as a market maker operating closer to the end of the trading period faces greater urgency in managing inventory imbalances. Since any remaining inventory at the end of the trading horizon incurs a penalty (\( \psi_0 > 0 \)), the market maker becomes **more sensitive to inventory fluctuations** and adjusts bid-ask spreads more aggressively to offset imbalances in the final trading steps.

While the general trends in the figures align with theoretical expectations, the visual representation of these effects may not be as pronounced due to the challenges associated with hyperparameter tuning in numerical experiments. Fine-tuning hyperparameters plays a crucial role in obtaining clear and visually striking results, and this process requires significant expertise in numerical optimization. The results presented here represent a best effort by the authors, but it is acknowledged that more refined parameter choices and computational techniques could yield more distinct and visually compelling patterns. Future work could focus on optimizing these numerical settings to enhance the clarity of the observed effects.

The following figures, \ref{fig:bid-time} and \ref{fig:ask-time}, illustrate the average behavior of the final policy under different trading times. \textit{Note:} Other state parameters differ from the previous settings. In these figures, the key variable that changes is the trading time, as it approaches the end of the trading period. However, for each trading time, the market maker maintains a fixed inventory of \textbf{100 shares} (i.e., the market maker has a \textbf{positive inventory}).  

As observed in the figures, the \textbf{bid spread decreases} (implying an increase in the bid price) as the trading period nears its end, while the \textbf{ask spread increases} (indicating a rise in the ask price). This behavior is intuitive, as the original value function imposes a \textbf{penalty on final inventory holdings}. Consequently, the market maker becomes more eager to offset its \textbf{existing inventory of 100 shares} before the trading period concludes.  

To achieve this, the market maker \textbf{lowers the bid price}, thereby reducing the likelihood of acquiring additional shares from the market. Simultaneously, it \textbf{raises the ask price}, increasing the probability of selling existing shares. These adjustments strategically help the market maker manage inventory risk while optimizing trading performance.

\begin{figure}[htbp]
    \centering
    \includegraphics[width=\linewidth]{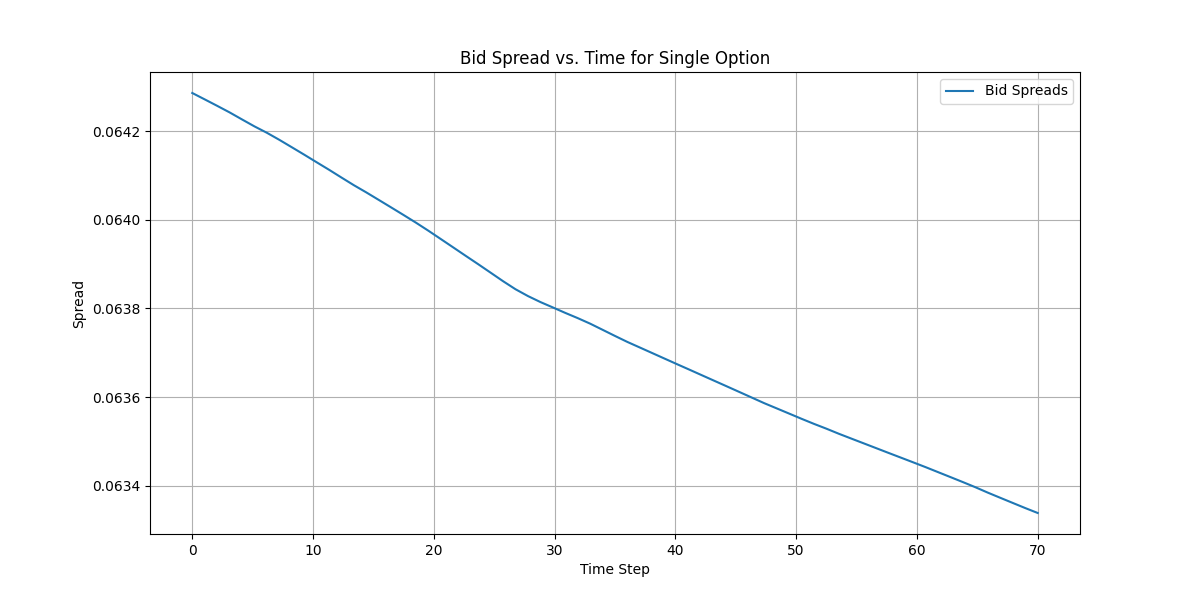}
    \caption{Bid spread over time. As the trading period nears its end, the bid spread decreases, implying an increase in bid price.}
    \label{fig:bid-time}
\end{figure}  

\begin{figure}[htbp]
    \centering
    \includegraphics[width=\linewidth]{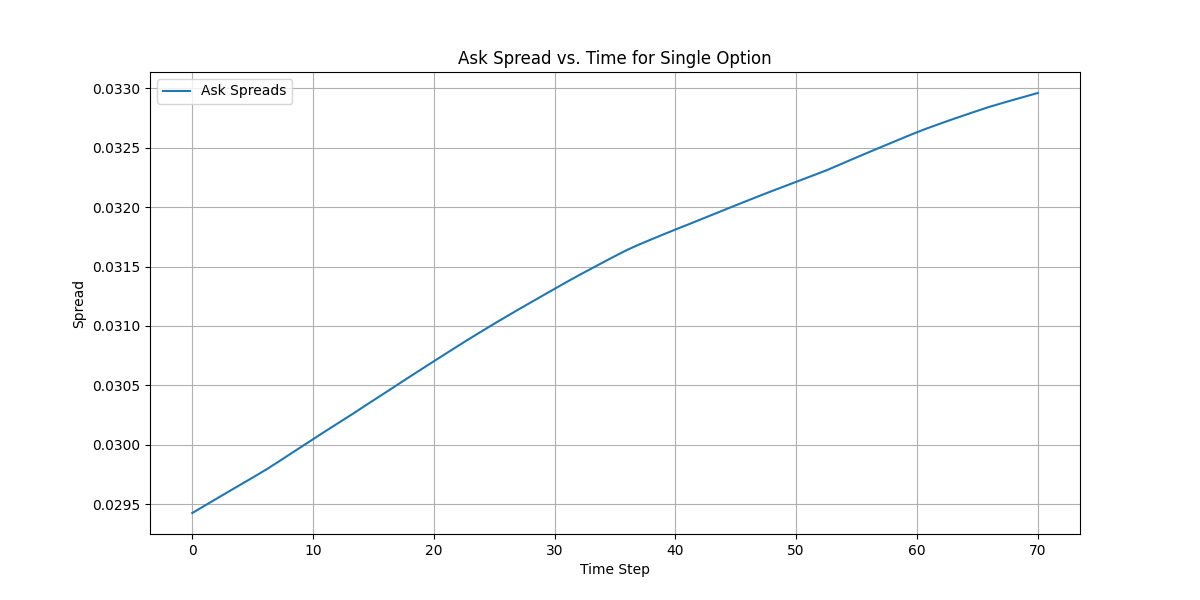}
    \caption{Ask spread over time. As the trading period nears its end, the ask spread increases, implying an increase in ask price.}
    \label{fig:ask-time}
\end{figure}

The following figure (\ref{fig:inventory-paths}) illustrates the inventory paths in the single-option case. To analyze inventory behavior, we generate hundreds of paths and compute both the mean and standard deviation.  

As shown in Figure \ref{fig:inventory-paths}, the market maker tends to accumulate inventory during the first half of the trading period, likely for speculative reasons or to take advantage of favorable pricing conditions. However, in the second half, the inventory level declines. This behavior aligns with intuition, as the penalty for holding inventory at the end of the trading period incentivizes the market maker to gradually reduce its holdings, thereby minimizing potential liquidation costs.  

\begin{figure}[htbp]
    \centering
    \includegraphics[width=0.8\linewidth]{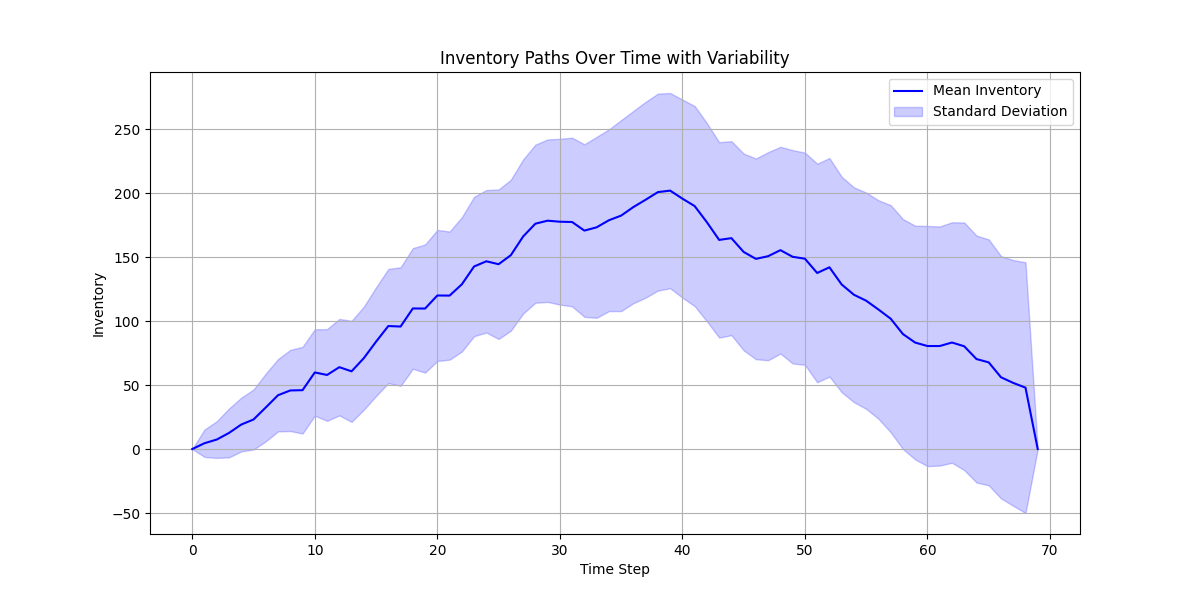}
    \caption{Inventory paths for the single-option case, displaying the mean and standard deviation.}
    \label{fig:inventory-paths}
\end{figure}

For the multi-dimensional option market-making case, we begin with the simplest two-option scenario as a sanity check. A key observation is that options with strikes further from the current price (out-of-the-money) tend to exhibit wider bid-ask spreads.  

\begin{figure}[htbp]
    \centering
    \includegraphics[width=\linewidth]{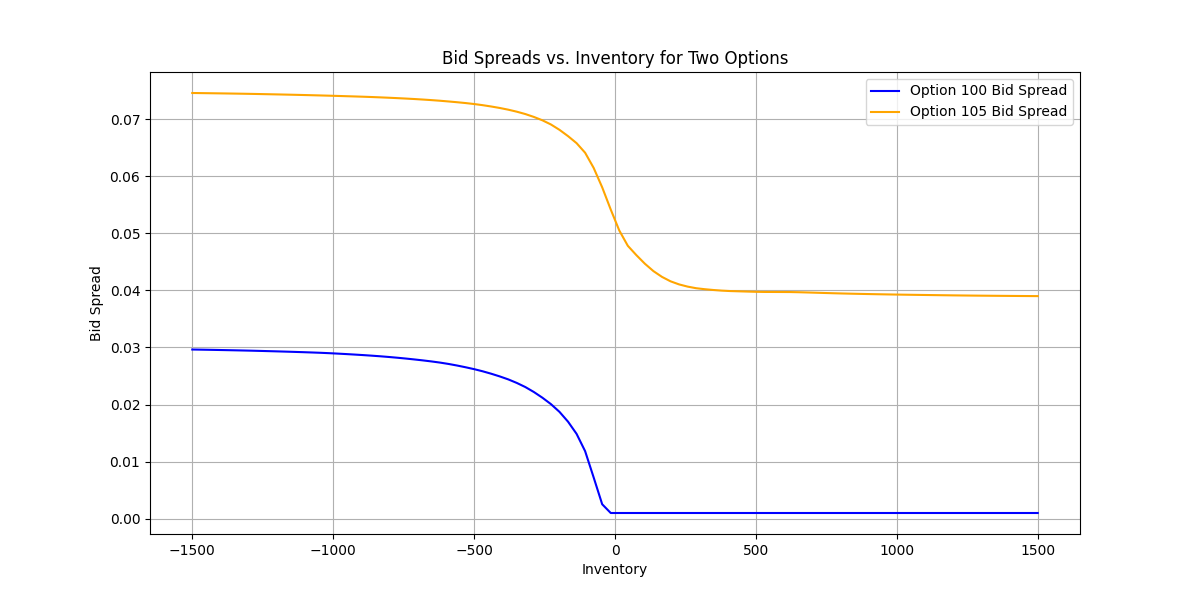}
    \caption{Bid spread as a function of inventory for the first option.}
    \label{fig:bid_spread_inv_option1}
\end{figure}

\begin{figure}[htbp]
    \centering
    \includegraphics[width=\linewidth]{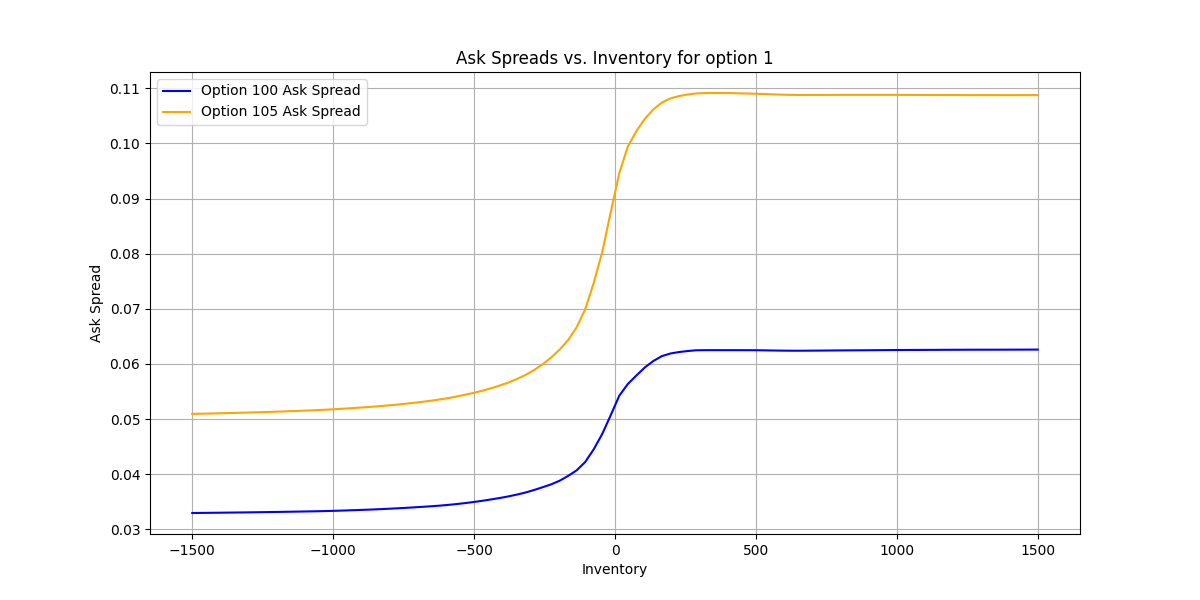}
    \caption{Ask spread as a function of inventory for the first option.}
    \label{fig:ask_spread_inv_option1}
\end{figure}

In Figures~\ref{fig:bid_spread_inv_option1} and~\ref{fig:ask_spread_inv_option1}, we examine the impact of inventory on bid-ask spreads by varying the inventory of the first option (with a lower strike price of $100$) while keeping the inventory of the second option (with a higher strike price of $105$) fixed at zero. Interestingly, increasing the inventory of the first option not only affects its own bid-ask spread but also influences the spread of the second option. This interdependence underscores the role of inventory risk across multiple options, highlighting the necessity of joint optimization in a multi-asset market-making strategy.  

Since this paper primarily focuses on the theoretical framework of reinforcement learning for market making, the numerical experiments serve as a validation step to ensure that the results align with general intuition.  

There are numerous potential improvements in the deep learning training process that could enhance stability, accuracy, and efficiency. However, these refinements are beyond the scope of this work and are left for future research.

\clearpage

\begin{thebibliography}{99}

\bibitem{muni2017modelling} Muni Toke, Ioane and Yoshida, Nakahiro.\newblock Modelling intensities of order flows in a limit order book.\newblock \emph{Quantitative Finance}, 17(5):683--701, 2017.

\bibitem{cartea2015order} Cartea, {\'A}lvaro and Jaimungal, Sebastian.\newblock Order-flow and liquidity provision.\newblock \emph{Available at SSRN 2553154}, 2015.

\bibitem{jia2022policy} Jia, Yanwei and Zhou, Xun Yu.\newblock Policy evaluation and temporal-difference learning in continuous time and space: A martingale approach.\newblock \emph{Journal of Machine Learning Research}, 23(154):1--55, 2022.

\bibitem{avellaneda2008high} Avellaneda, Marco and Stoikov, Sasha.\newblock High-frequency trading in a limit order book.\newblock \emph{Quantitative Finance}, 8(3):217--224, 2008.

\bibitem{stoikov2009option} Stoikov, Sasha and Sa{\u{g}}lam, Mehmet.\newblock Option market making under inventory risk.\newblock \emph{Review of Derivatives Research}, 12:55--79, 2009.

\bibitem{grossman1988liquidity} Grossman, Sanford J and Miller, Merton H.\newblock Liquidity and market structure.\newblock \emph{The Journal of Finance}, 43(3):617--633, 1988.

\bibitem{cartea2020market} Cartea, {\'A}lvaro and Wang, Yixuan.\newblock Market making with alpha signals.\newblock \emph{International Journal of Theoretical and Applied Finance}, 23(03):2050016, 2020.

\bibitem{cartea2014buy} Cartea, {\'A}lvaro and Jaimungal, Sebastian and Ricci, Jason.\newblock Buy low, sell high: A high-frequency trading perspective.\newblock \emph{SIAM Journal on Financial Mathematics}, 5(1):415--444, 2014.

\bibitem{cartea2017algorithmic} Cartea, {\'A}lvaro and Donnelly, Ryan and Jaimungal, Sebastian.\newblock Algorithmic trading with model uncertainty.\newblock \emph{SIAM Journal on Financial Mathematics}, 8(1):635--671, 2017.

\bibitem{baldacci2021algorithmic} Baldacci, Bastien and Bergault, Philippe and Gu{\'e}ant, Olivier.\newblock Algorithmic market making for options.\newblock \emph{Quantitative Finance}, 21(1):85--97, 2021.

\bibitem{bergault2021closed} Bergault, Philippe and Evangelista, David and Gu{\'e}ant, Olivier and Vieira, Douglas.\newblock Closed-form approximations in multi-asset market making.\newblock \emph{Applied Mathematical Finance}, 28(2):101--142, 2021.

\bibitem{wang2020continuous} Wang, Haoran and Zhou, Xun Yu.\newblock Continuous-time mean--variance portfolio selection: A reinforcement learning framework.\newblock \emph{Mathematical Finance}, 30(4):1273--1308, 2020.

\bibitem{ho1981optimal} Ho, Thomas and Stoll, Hans R.\newblock Optimal dealer pricing under transactions and return uncertainty.\newblock \emph{Journal of Financial Economics}, 9(1):47--73, 1981.

\bibitem{beysolow2019market} Beysolow II, Taweh.\newblock Market making via reinforcement learning.\newblock \emph{Applied Reinforcement Learning with Python: With OpenAI Gym, Tensorflow, and Keras}, pages 77--94, 2019.

\bibitem{spooner2020robust} Spooner, Thomas and Savani, Rahul.\newblock Robust market making via adversarial reinforcement learning.\newblock \emph{arXiv preprint arXiv:2003.01820}, 2020.

\bibitem{ganesh2019reinforcement} Ganesh, Sumitra and Vadori, Nelson and Xu, Mengda and Zheng, Hua and Reddy, Prashant and Veloso, Manuela.\newblock Reinforcement learning for market making in a multi-agent dealer market.\newblock \emph{arXiv preprint arXiv:1911.05892}, 2019.

\bibitem{sadighian2020extending} Sadighian, Jonathan.\newblock Extending deep reinforcement learning frameworks in cryptocurrency market making.\newblock \emph{arXiv preprint arXiv:2004.06985}, 2020.

\bibitem{hu2013kullback} Hu, Zhaolin and Hong, L Jeff.\newblock Kullback-Leibler divergence constrained distributionally robust optimization.\newblock \emph{Available at Optimization Online}, 1(2):9, 2013.

\bibitem{jiang2016data} Jiang, Ruiwei and Guan, Yongpei.\newblock Data-driven chance constrained stochastic program.\newblock \emph{Mathematical Programming}, 158(1-2):291--327, 2016.

\bibitem{esfahani2015data} Esfahani, Peyman Mohajerin and Kuhn, Daniel.\newblock Data-driven distributionally robust optimization using the Wasserstein metric: Performance guarantees and tractable reformulations.\newblock \emph{arXiv preprint arXiv:1505.05116}, 2015.

\bibitem{gao2022wasserstein} Gao, Rui and Chen, Xi and Kleywegt, Anton J.\newblock Wasserstein distributionally robust optimization and variation regularization.\newblock \emph{Operations Research}, 2022.

\bibitem{michaud1989markowitz} Michaud, Richard O.\newblock The Markowitz optimization enigma: Is ‘optimized’optimal?\newblock \emph{Financial Analysts Journal}, 45(1):31--42, 1989.

\bibitem{abdullah2019wasserstein} Abdullah, Mohammed Amin and Ren, Hang and Ammar, Haitham Bou and Milenkovic, Vladimir and Luo, Rui and Zhang, Mingtian and Wang, Jun.\newblock Wasserstein robust reinforcement learning.\newblock \emph{arXiv preprint arXiv:1907.13196}, 2019.

\bibitem{hou2020robust} Hou, Linfang and Pang, Liang and Hong, Xin and Lan, Yanyan and Ma, Zhiming and Yin, Dawei.\newblock Robust reinforcement learning with Wasserstein constraint.\newblock \emph{arXiv preprint arXiv:2006.00945}, 2020.

\bibitem{kyle1985continuous} Kyle, Albert S.\newblock Continuous auctions and insider trading.\newblock \emph{Econometrica: Journal of the Econometric Society}, pages 1315--1335, 1985.

\end{thebibliography}

\clearpage
\appendix

\section{Derivation of HJB equation under Deterministic Policy} \label{HJB deterministic}
According to the definition of value function under deterministic policy is
\begin{align}
    &V_t(\boldsymbol{Q}_t, \boldsymbol{P}_t, S_t) \nonumber \\ 
    = \underset{(\epsilon^-_u,\epsilon^+_u): t\leq u\leq T}{\max} &\mathbb{E} \Bigg[   \int_t^T \Big[(\boldsymbol{\epsilon}_u^+)^{T} d\boldsymbol N_u^+ + (\boldsymbol{\epsilon}_u^-)^{T} d\boldsymbol N_u^-  \Big] du + \int_t^T \boldsymbol{Q}_u^{T} \Big( \frac{\partial \boldsymbol p_u}{\partial u} +  \frac{\partial \boldsymbol p_u^2}{\partial S_u^2} \Big) du \nonumber \\
    &- \psi_0 \sum_{i = 1}^n  (P_{iT} Q_{iT})^2   \hspace{0.05cm} \Big| \hspace{0.05cm} \boldsymbol{Q}_t, \boldsymbol{P}_t, S_t  \Bigg] \nonumber \\
    = \underset{(\epsilon^-_s,\epsilon^+_s): t\leq s\leq T}{\max} &\mathbb{E} \Bigg[   \int_t^{t + \Delta t} \Big[(\boldsymbol{\epsilon}_u^+)^{T} d\boldsymbol N_u^+ + (\boldsymbol{\epsilon}_u^-)^{T} d\boldsymbol N_u^-  \Big] du + \int_t^{t + 
    \Delta t} \boldsymbol{Q}_u^{T} \Big( \frac{\partial \boldsymbol p_u}{\partial u} +  \frac{\partial \boldsymbol p_u^2}{\partial S_u^2} \Big) du \nonumber \\
    & + V_{t + \Delta t}(\boldsymbol{Q}_t + \Delta \boldsymbol{Q}, \boldsymbol{P}_t + \Delta \boldsymbol{P}, S_t + \Delta S) \hspace{0.05cm} \Big| \hspace{0.05cm}  \boldsymbol{Q}_t, \boldsymbol{P}_t, S_t  \Bigg] 
\end{align}

For the last term, $V_{t + \Delta t}(\boldsymbol{Q}_t + \Delta \boldsymbol{Q}, \boldsymbol{P}_t + \Delta \boldsymbol{P}, S_t + \Delta S)$, the following equality holds by Ito calculus
\begin{align}
     &V_{t + \Delta t}(\boldsymbol{Q}_t + \Delta \boldsymbol{Q}, \boldsymbol{P}_t + \Delta \boldsymbol{P}, S_t + \Delta S) \nonumber \\
    = \ & V_{t + \Delta t}(\boldsymbol{Q}_t, \boldsymbol{P}_t + \Delta \boldsymbol{P}, S_t + \Delta S) \prod_{i = 1}^{n}(1 - \Delta N_{it}^+)(1 - \Delta N_{it}^-) \nonumber \\
    +& \sum_{i = 1}^{n} \Big[V_{t + \Delta t}(\boldsymbol{Q}_t + \boldsymbol{\Delta}_i, \boldsymbol{P}_t + \Delta \boldsymbol{P}, S_t + \Delta S) \Delta N_{it}^+ + V_{t + \Delta t}(\boldsymbol{Q}_t - \boldsymbol{\Delta}_i, \boldsymbol{P}_t + \Delta \boldsymbol{P}, S_t + \Delta S) \Delta N_{it}^-\Big] \nonumber \\
    = \ & \bigg[V_t(\boldsymbol{Q}_t, \boldsymbol{P}_t, S_t) + \partial_t V_t(\boldsymbol{Q}_t, \boldsymbol{P}_t, S_t) \Delta t + \sigma S_t \partial_S V_t( \boldsymbol{Q}_t, \boldsymbol{P}_t, S_t) \Delta W_t^s + \frac{\sigma^2 S_t^2}{2} \partial_{SS} V_t(\boldsymbol{Q}_t, \boldsymbol{P}_t, S_t) \Delta t \nonumber \\
    +& \nabla_{\boldsymbol{P}} V_t(\boldsymbol{Q}_t, \boldsymbol{P}_t, S_t) \Delta \boldsymbol{P} + \frac{1}{2}(\Delta \boldsymbol{P})^{T}  \boldsymbol{H}_{\boldsymbol{P}} \Delta \boldsymbol{P} \bigg]  \prod_{i = 1}^{n} (1 - \Delta N_{it}^+)(1 - \Delta N_{it}^-) \nonumber \\ 
    + & \sum_{i = 1}^{n} \Big[V_{t + \Delta t}(\boldsymbol{Q}_t + \boldsymbol{\Delta}_i, \boldsymbol{P}_t + \Delta \boldsymbol{P}, S_t + \Delta S) \Delta N_{it}^+ + V_{t + \Delta t}(\boldsymbol{Q}_t - \boldsymbol{\Delta}_i, \boldsymbol{P}_t + \Delta \boldsymbol{P}, S_t + \Delta S) \Delta N_{it}^-\Big]
\end{align}

By taking the expectation conditioned on state at time $t$,
\begin{align}
    & \mathbb{E} \bigg[ V_{t + \Delta t}(\boldsymbol{Q_t} + \Delta \boldsymbol{Q}, \boldsymbol{P}_t + \Delta \boldsymbol{P}, S_t + \Delta S) \hspace{0.05cm} \Big | \hspace{0.05cm} \boldsymbol{Q_t}, \boldsymbol{P}_t, S_t \bigg] \nonumber \\
    = & \  V_t(\boldsymbol{Q}_t, \boldsymbol{P}_t, S_t) + \bigg[\partial_t V_t(\boldsymbol{Q}_t, \boldsymbol{P}_t, S_t) + \frac{\sigma^2 S_t^2}{2} \partial_{SS} V_t(\boldsymbol{Q}_t, \boldsymbol{P}_t, S_t) - \sum_{i = 1}^{n} (\lambda_{it}^+ + \lambda_{it}^-) V_t(\boldsymbol{Q}_t, \boldsymbol{P}_t, S_t) \bigg] \Delta t  \nonumber \\
    + & \bigg[ \sum_{i = 1}^{n}  V_t(\boldsymbol{Q}_t + \boldsymbol{\Delta}_i, \boldsymbol{P}_t, S_t) \lambda_{it}^+ + V_t(\boldsymbol{Q}_t - \boldsymbol{\Delta}_i, \boldsymbol{P}_t, S_t) \lambda_{it}^-  \bigg] \Delta t  \nonumber \\
    +&  \  \mathbb{E} \bigg[    \nabla_{\boldsymbol{P}} V_t \Delta \boldsymbol{P} + \frac{1}{2}(\Delta \boldsymbol{P})^{T}  \boldsymbol{H}_{\boldsymbol{P}} \Delta \boldsymbol{P} \hspace{0.05cm}\bigg| \hspace{0.05cm} \boldsymbol{Q}_t, \boldsymbol{P}_t, S_t \bigg] \nonumber \\
    =  & \ \ V_t(\boldsymbol{Q}_t, \boldsymbol{P}_t, S_t) + \bigg[\partial_t V_t(\boldsymbol{Q}_t, \boldsymbol{P}_t, S_t) + \frac{\sigma^2 S_t^2}{2} \partial_{SS} V_t(\boldsymbol{Q}_t, \boldsymbol{P}_t, S_t) - \sum_{i = 1}^{n} (\lambda_{it}^+ + \lambda_{it}^-) V_t(\boldsymbol{Q}_t, \boldsymbol{P}_t, S_t) \bigg] \Delta t  \nonumber \\
    +& \bigg[ \sum_{i = 1}^{n}  V_t(\boldsymbol{Q}_t + \boldsymbol{\Delta}_i, \boldsymbol{P}_t, S_t) \lambda_{it}^+ + V_t(\boldsymbol{Q}_t - \boldsymbol{\Delta}_i, \boldsymbol{P}_t, S_t) \lambda_{it}^-  \bigg] \Delta t  \nonumber \\
    + & \  (\nabla_{\boldsymbol{P}} V_t)^{T} \big( \frac{\partial \boldsymbol{p}_t}{\partial t}  + \frac{\sigma^2 S_t^2}{2} \frac{\partial^2 \boldsymbol{p}_t}{\partial S_t^2} \big) \Delta t + \Big[\frac{\sigma^2 S_t^2}{2} (\frac{\partial \boldsymbol{p}_t}{\partial S_t})^T \boldsymbol{H}_{\boldsymbol{P}}   (\frac{\partial \boldsymbol{p}_t}{\partial S_t}) + \frac{1}{2} \textbf{Tr} (\boldsymbol{V}_t^{T} \boldsymbol{H}_{\boldsymbol{P}} \boldsymbol{V}_t) \Big] \Delta t
\end{align}

After plugging the conditional expectation of  $V_{t + \Delta t}(\boldsymbol{Q}_t + \Delta \boldsymbol{Q}, \boldsymbol{P}_t + \Delta \boldsymbol{P}, S_t + \Delta S)$, and cancel the time step $\Delta t$, the HJB equation under deterministic yields

\begin{align}
    &\max_{\boldsymbol \epsilon_t^+, \boldsymbol \epsilon_t^- \in \mathbb R^{n}_+} \Bigg\{  \sum_{i = 1}^{n} \lambda_{it} \big(  \epsilon_{it}^+ + \Delta_i^+ V_t(\boldsymbol{Q}_t + \boldsymbol \Delta_i, \boldsymbol{P}_t, S_t)  \big) + \sum_{i = 1}^{n} \lambda_{it} \big( \epsilon_{it}^- + \Delta_i^- V_t(\boldsymbol{Q}_t - \boldsymbol \Delta_i, \boldsymbol{P}_t, S_t) \big) \Bigg\} \nonumber \\
    &+ \boldsymbol{Q}_t^{T} \big( 
    \frac{\partial \boldsymbol{p_t}}{\partial t} + \frac{\partial^2 \boldsymbol p_t}{\partial S_t^2}\big) + \partial_t V_t( \boldsymbol{Q}_t, \boldsymbol{P}_t, S_t) + \frac{\sigma^2 S_t^2}{2}\partial_{SS} V_t( \boldsymbol{Q_t}, \boldsymbol{P}_t, S_t) \nonumber \\
    &+ (\nabla_{\boldsymbol{P}} V_t)^{T} \big( \frac{\partial \boldsymbol{p}_t}{\partial t}  + \frac{\sigma^2S_t^2}{2} \frac{\partial^2 \boldsymbol{p}_t}{\partial S_t^2} \big) + \Big[\frac{\sigma^2 S_t^2}{2} (\frac{\partial \boldsymbol{p}_t}{\partial S_t})^T \boldsymbol{H}_{\boldsymbol{P}}   (\frac{\partial \boldsymbol{p}_t}{\partial S_t}) + \frac{1}{2} \textbf{Tr} (\boldsymbol{V}_t^{T} \boldsymbol{H}_{\boldsymbol{P}} \boldsymbol{V}_t) \Big] = 0
\end{align}

\section{Concavity of Maximization Bracket}\label{concavity of maximization}
Considering the functional inside the maximization problem, our goal here to show that it has the global maximizer inside the convex set of measures

First, define the functional as 
\begin{align*}
    f(\boldsymbol{\pi})
    = &\int_{\boldsymbol{\epsilon}^+} \boldsymbol{\pi}(\boldsymbol{\epsilon_t^+} | \  t, \boldsymbol{Q}_t, \boldsymbol{P}_t, S_t)  \sum_{i = 1}^{n} \lambda_{it} \big(  \epsilon_{it}^+ + V_t(\boldsymbol{Q}_t + \boldsymbol \Delta_i, \boldsymbol{P}_t, S_t) - V_t(\boldsymbol{Q}_t, \boldsymbol{P}_t, S_t) \big) d\boldsymbol{\epsilon_t^+} \\
    + & \int_{\boldsymbol{\epsilon}^-} \boldsymbol{\pi}(\boldsymbol{\epsilon_t^-} | \ t, \boldsymbol{Q}_t, \boldsymbol{P}_t, S_t)  \sum_{i = 1}^{n} \lambda_{it} \big( \epsilon_{it}^- + V_t(\boldsymbol{Q}_t - \boldsymbol \Delta_i, \boldsymbol{P}_t, S_t) - V_t(\boldsymbol{Q}_t, \boldsymbol{P}_t, S_t)\big) d\boldsymbol{\epsilon_t^-} \\
    - & \gamma \int_{\boldsymbol{\epsilon}} \boldsymbol{\pi}(\boldsymbol{\epsilon}_t^{\pm} | \boldsymbol{Q}_t, \boldsymbol{P}_t, S_t) \log \boldsymbol{\pi}(\boldsymbol{\epsilon}_t^{\pm} | \boldsymbol{Q}_t, \boldsymbol{P}_t, S_t) d\boldsymbol{\epsilon}
\end{align*}

If there are two polcies $\boldsymbol{\pi}_1$ and $\boldsymbol{\pi}_2$, and let $p \in (0, 1)$, there is the following derivation,
\begin{align}
    &f(p\boldsymbol{\pi}_1 + (1 - p) \boldsymbol{\pi}_2) \nonumber \\
    = & \int_{\boldsymbol{\epsilon}} \big( p\boldsymbol{\pi}_1 + (1 - p) \boldsymbol{\pi}_2 \big)  \sum_{i = 1}^{n} \lambda_{it} \big(  \epsilon_{it}^+ + V_t(\boldsymbol{Q}_t + \boldsymbol \Delta_i, \boldsymbol{P}_t, S_t) - V_t(\boldsymbol{Q}_t, \boldsymbol{P}_t, S_t) \big) d\boldsymbol{\epsilon_t^+} \nonumber \\
    + & \int_{\boldsymbol{\epsilon}} \big( p\boldsymbol{\pi}_1 + (1 - p) \boldsymbol{\pi}_2 \big)  \sum_{i = 1}^{n} \lambda_{it} \big( \epsilon_{it}^- + V_t(\boldsymbol{Q}_t - \boldsymbol \Delta_i, \boldsymbol{P}_t, S_t) - V_t(\boldsymbol{Q}_t, \boldsymbol{P}_t, S_t)\big) d\boldsymbol{\epsilon_t^-} \nonumber \\
    - & \  \gamma \int_{\boldsymbol{\epsilon}}  \big( p\boldsymbol{\pi}_1 + (1 - p) \boldsymbol{\pi}_2 \big) \log  \big( p\boldsymbol{\pi}_1 + (1 - p) \boldsymbol{\pi}_2 \big) d\boldsymbol{\epsilon} \nonumber \\
    \leq & \int_{\boldsymbol{\epsilon}} \big( p\boldsymbol{\pi}_1 + (1 - p) \boldsymbol{\pi}_2 \big)  \sum_{i = 1}^{n} \lambda_{it} \big(  \epsilon_{it}^+ + V_t(\boldsymbol{Q}_t + \boldsymbol \Delta_i, \boldsymbol{P}_t, S_t) - V_t(\boldsymbol{Q}_t, \boldsymbol{P}_t, S_t) \big) d\boldsymbol{\epsilon_t^+} \nonumber \\
    + & \int_{\boldsymbol{\epsilon}} \big( p\boldsymbol{\pi}_1 + (1 - p) \boldsymbol{\pi}_2 \big)  \sum_{i = 1}^{n} \lambda_{it} \big( \epsilon_{it}^- + V_t(\boldsymbol{Q}_t - \boldsymbol \Delta_i, \boldsymbol{P}_t, S_t) - V_t(\boldsymbol{Q}_t, \boldsymbol{P}_t, S_t)\big) d\boldsymbol{\epsilon_t^-} \nonumber \\
    - & \ \gamma \Big( p \int_{\boldsymbol{\epsilon}} \boldsymbol{\pi}_1 \log  \boldsymbol{\pi}_1 d\boldsymbol{\epsilon} + (1 - p) \int_{\boldsymbol{\epsilon}} \boldsymbol{\pi}_2 \log  \boldsymbol{\pi}_2 d \boldsymbol{\epsilon} \Big) \nonumber \\
    = & \  p f(\boldsymbol{\pi}_1) + (1 - p) f(\boldsymbol{\pi}_2)
\end{align}

Since $x \log x$ is a concave function of $x$, this results the above inequality. 

Therefore, we know that the maximizer is an interior point, and we can use the calculus of variation techniques to get the optimal policy.

\section{Derivation of Optimal Policy}\label{Derive Optimal Policy}

Based on the above section's argument about the convexity of the maximization problem, we can know that the local maximizer is the global maximizer, and the global maximum is the interior point of all admissible probability measures. Therefore, we can apply a trick from the calculus of variation to obtain the optimal policy.

Let $\boldsymbol \pi^*$ be the optimal policy, and perturbed the policy by adding $\delta \boldsymbol{\pi}$, which will yields the following, 

\begin{align}
    0 &= \int_{\boldsymbol{\epsilon}^+} \delta \boldsymbol{\pi}  \sum_{i = 1}^{n} \lambda_{it} \big(  \epsilon_{it}^+ + \Delta_i^+ V_t(\boldsymbol{Q}_t + \boldsymbol \Delta_i, \boldsymbol{P}_t, S_t) \big) d\boldsymbol{\epsilon_t^+} \nonumber \\
    &+ \int_{\boldsymbol{\epsilon}^-} \delta \boldsymbol{\pi}  \sum_{i = 1}^{n} \lambda_{it} \big( \epsilon_{it}^- + \Delta_i^- V_t(\boldsymbol{Q}_t - \boldsymbol \Delta_i, \boldsymbol{P}_t, S_t)\big) d\boldsymbol{\epsilon_t^-} \nonumber \\
    &- \gamma \int_{\boldsymbol{\epsilon_t}} \boldsymbol \pi^* \frac{\delta \boldsymbol \pi}{\boldsymbol \pi^*} d\boldsymbol{\epsilon_t} - \gamma  \int_{\boldsymbol{\epsilon_t}} \delta \boldsymbol \pi \log \boldsymbol  \pi^* d\boldsymbol{\epsilon_t}
\end{align}

In addition, $\delta \boldsymbol{\pi}$ has the following constrain, 
\begin{align}
    \int_{\boldsymbol{\epsilon_t}} \delta \boldsymbol{\pi} d\epsilon_t = 0
\end{align}

From the above two equations, there are
\begin{align}
    C &=  \sum_{i = 1}^{n} \lambda_{it} \big(  \epsilon_{it}^+ + \Delta_i^+ V_t(\boldsymbol{Q}_t + \boldsymbol \Delta_i, \boldsymbol{P}_t, S_t)  \big) + \sum_{i = 1}^{n} \lambda_{it} \big( \epsilon_{it}^- + \Delta_i^- V_t(\boldsymbol{Q}_t - \boldsymbol \Delta_i, \boldsymbol{P}_t, S_t)\big) \nonumber \\
    &- \gamma \log \boldsymbol{\pi}^*(\boldsymbol{\epsilon_t}  | \boldsymbol{Q}_t, \boldsymbol{P}_t, S_t)
\end{align}

We could get the following derivation
\begin{align}
    \boldsymbol{\pi}^* 
    \propto & \exp \Big \{ \frac{1}{\gamma} \sum_{i = 1}^{n} \lambda_{it} \big(  \epsilon_{it}^+ + \Delta_i^+ V_t(\boldsymbol{Q}_t, \boldsymbol{P}_t, S_t) \big) + \sum_{i = 1}^{n} \lambda_{it} \big( \epsilon_{it}^- + \Delta_i^- V_t(\boldsymbol{Q}_t, \boldsymbol{P}_t, S_t)\big)    \Big\} \nonumber \\
    =& \frac{\exp \Big \{ \frac{1}{\gamma} \sum_{i = 1}^{n} \lambda_{it} \big(  \epsilon_{it}^+ + \Delta_i^+ V_t(\boldsymbol{Q}_t, \boldsymbol{P}_t, S_t) \big) + \sum_{i = 1}^{n} \lambda_{it} \big( \epsilon_{it}^- + \Delta_i^- V_t(\boldsymbol{Q}_t, \boldsymbol{P}_t, S_t)\big)    \Big\}}{\int_{\boldsymbol{\epsilon_t}}\exp \Big \{ \frac{1}{\gamma} \sum_{i = 1}^{n} \lambda_{it} \big(  \epsilon_{it}^+ + \Delta_i^+ V_t(\boldsymbol{Q}_t, \boldsymbol{P}_t, S_t) \big) + \sum_{i = 1}^{n} \lambda_{it} \big( \epsilon_{it}^- + \Delta_i^- V_t(\boldsymbol{Q}_t, \boldsymbol{P}_t, S_t)\big)    \Big\} d\boldsymbol{\epsilon_t}} \nonumber 
\end{align}
where, 
\begin{align}
    \Delta_i^+ V_t(\boldsymbol{Q}_t, \boldsymbol {P}_t, S_t) &= V_t(\boldsymbol{Q}_t + \boldsymbol{\Delta}_i, \boldsymbol{P}_t, S_t) - V_t(\boldsymbol{Q}_t, \boldsymbol{P}_t, S_t), \\
    \Delta_i^- V_t(\boldsymbol{Q}_t, \boldsymbol{P}_t, S_t) &= V_t(\boldsymbol{Q}_t - \boldsymbol{\Delta}_i, \boldsymbol{P}_t, S_t) - V_t(\boldsymbol{Q}_t, \boldsymbol{P}_t, S_t).
\end{align}

\section{Proof of Policy Improvement Theorem}\label{Policy Improvement}
 Let's consider the following case, from time period $[t, s]$, one uses $\widetilde{\boldsymbol{\pi}}$, and uses $\boldsymbol{\pi}$ for time $[s, T]$. 
    By the Ito formula, the following equation holds, (notice that $\boldsymbol{Q}_u^{\widetilde{\boldsymbol{\pi}}}$ means the inventory processes under policy $\widetilde{\boldsymbol{\pi}}$)
    \begin{align}
        & V_t^{\boldsymbol{\pi}}(\boldsymbol{Q}_t, \boldsymbol{P}_t, S_t) \nonumber \\ 
        = \ & \mathbb E \Big[ V^{\boldsymbol{\pi}}_s(\boldsymbol{Q}^{\widetilde{\boldsymbol{\pi}}}_s, \boldsymbol{P}_s, S_s) \nonumber \\
        -& \int_t^s \int_{\boldsymbol{\epsilon_u}} \widetilde{\boldsymbol{\pi}} (\boldsymbol{\epsilon_u}) \sum_{i = 1}^N \Delta_i^+ V_u^{\boldsymbol{\pi}}(\boldsymbol{Q}_u^{\widetilde{\boldsymbol{\pi}}}, \boldsymbol {P}_u, S_u) \Delta N_{iu}^+ d\boldsymbol{\epsilon_u} + \Delta_i^- V_u^{\boldsymbol{\pi}}(\boldsymbol{Q}_u^{\widetilde{\boldsymbol{\pi}}}, \boldsymbol {P}_u, S_u) \Delta N_{iu}^- d\boldsymbol{\epsilon_u}  \nonumber \\ 
        -& \int_t^s \bigg( \partial_t V_u^{\boldsymbol{\pi}}( \boldsymbol{Q}_u^{\widetilde{\boldsymbol{\pi}}}, \boldsymbol{P}_u, S_u) + \frac{\sigma^2 S_u^2}{2}\partial_{SS} V_u^{\boldsymbol{\pi}}(\boldsymbol{Q}_u^{\widetilde{\boldsymbol{\pi}}}, \boldsymbol{P}_u, S_u) + (\nabla_{\boldsymbol{P}} V_u^{\boldsymbol{\pi}})^{T} \big( \frac{\partial \boldsymbol{p}_u}{\partial t}  + \frac{\sigma^2S_u^2}{2} \frac{\partial^2 \boldsymbol{p}_u}{\partial S^2} \big) \nonumber \\ 
        +& \Big[\frac{\sigma^2 S_u^2}{2} (\frac{\partial \boldsymbol{p}_u}{\partial S_t})^T \boldsymbol{H}^{\boldsymbol{\pi}}_{\boldsymbol{P}}   (\frac{\partial \boldsymbol{p}_u}{\partial S_t}) + \frac{1}{2} \textbf{Tr} (\boldsymbol{V}_u^{T} \boldsymbol{H}^{\boldsymbol{\pi}}_{\boldsymbol{P}} \boldsymbol{V}_u) \Big] \bigg) du \  \Big | \ t, \boldsymbol{Q}_t, \boldsymbol{P}_t, S_t\Big]  \nonumber \\
        = \ & \mathbb E  \Big[ V^{\boldsymbol{\pi}}_s(\boldsymbol{Q}^{\widetilde{\boldsymbol{\pi}}}_s, \boldsymbol{P}_s, S_s) \ \Big | \ t, \boldsymbol{Q}_t, \boldsymbol{P}_t, S_t \Big ] \nonumber \\
        -& \int_t^s \int_{\boldsymbol{\epsilon_u}} \widetilde{\boldsymbol{\pi}} (\boldsymbol{\epsilon_u}) \sum_{i = 1}^N \lambda_{iu}^+ \Delta_i^+ V_u^{\boldsymbol{\pi}}(\boldsymbol{Q}_u^{\widetilde{\boldsymbol{\pi}}}, \boldsymbol {P}_u, S_u) d\boldsymbol{\epsilon_u} + \lambda_{iu}^- \Delta_i^- V_u^{\boldsymbol{\pi}}(\boldsymbol{Q}_u^{\widetilde{\boldsymbol{\pi}}}, \boldsymbol {P}_u, S_u)  d\boldsymbol{\epsilon_u}  \nonumber \\ 
         -& \int_t^s \bigg( \partial_t V_u^{\boldsymbol{\pi}}( \boldsymbol{Q}_u^{\widetilde{\boldsymbol{\pi}}}, \boldsymbol{P}_u, S_u) + \frac{\sigma^2 S_u^2}{2}\partial_{SS} V_u^{\boldsymbol{\pi}}(\boldsymbol{Q}_u^{\widetilde{\boldsymbol{\pi}}}, \boldsymbol{P}_u, S_u) + (\nabla_{\boldsymbol{P}} V_u^{\boldsymbol{\pi}})^{T} \big( \frac{\partial \boldsymbol{p}_u}{\partial t}  + \frac{\sigma^2 S_u^2}{2} \frac{\partial^2 \boldsymbol{p}_u}{\partial S^2} \big) \nonumber \\ 
         +& \Big[\frac{\sigma^2 S_u^2}{2} (\frac{\partial \boldsymbol{p}_u}{\partial S_t})^T \boldsymbol{H}^{\boldsymbol{\pi}}_{\boldsymbol{P}}   (\frac{\partial \boldsymbol{p}_u}{\partial S_t}) + \frac{1}{2} \textbf{Tr} (\boldsymbol{V}_u^{T} \boldsymbol{H}^{\boldsymbol{\pi}}_{\boldsymbol{P}} \boldsymbol{V}_u) \Big] \bigg) du 
    \end{align}

    In addition, we know that the following equation
    \begin{align}
    & \int_{\boldsymbol{\epsilon}}{\boldsymbol{\pi}}(\boldsymbol{\epsilon})  \sum_{i = 1}^{n} \Big( \lambda_{it}^+ \big(  \epsilon_{it}^+ + \Delta_i^+ V_t^{\boldsymbol{\pi}}(\boldsymbol{Q}_t, \boldsymbol{P}_t, S_t) \big) + \lambda_{it}^- \big( \epsilon_{it}^- + \Delta_i^- V_t^{\boldsymbol{\pi}}(\boldsymbol{Q}_t, \boldsymbol{P}_t, S_t) \Big) d\boldsymbol{\epsilon} \nonumber \\
    -& \ \gamma \int_{\boldsymbol{\epsilon}} {\boldsymbol{\pi}}(\boldsymbol{\epsilon}_t) \log {\boldsymbol{\pi}}(\boldsymbol{\epsilon}_t) d\boldsymbol{\epsilon}_t +  \boldsymbol{Q}_t^{T} \big( \frac{\partial \boldsymbol{p_t}}{\partial t} + \frac{\partial^2 \boldsymbol p_t}{\partial S_t^2}\big) + \partial_t V_t^{\boldsymbol{\pi}}( \boldsymbol{Q}_t, \boldsymbol{P}_t, S_t) + \frac{\sigma^2 S_t^2}{2}\partial_{SS} V_t^{\boldsymbol{\pi}}( \boldsymbol{Q_t}, \boldsymbol{P}_t, S_t) \nonumber \\
    +& \ (\nabla_{\boldsymbol{P}} V_t^{\boldsymbol{\pi}})^{T} \big( \frac{\partial \boldsymbol{p}_t}{\partial t}  + \frac{\sigma^2S_t^2}{2} \frac{\partial^2 \boldsymbol{p}_t}{\partial S_t^2} \big) + \Big[\frac{\sigma_t^2 S_t^2}{2} (\frac{\partial \boldsymbol{p}_t}{\partial S_t})^T \boldsymbol{H}^{\boldsymbol{\pi}}_{\boldsymbol{P}}   (\frac{\partial \boldsymbol{p}_t}{\partial S_t}) + \frac{1}{2} \textbf{Tr} (\boldsymbol{V}_t^{T} \boldsymbol{H}^{\boldsymbol{\pi}}_{\boldsymbol{P}} \boldsymbol{V}_t) \Big] = 0
    \end{align}
    
    By the construction of the $\widetilde{\boldsymbol{\pi}}$, it is the maximizer of the following optimization question,

    \begin{align}
        \max_{\widetilde{\boldsymbol{\pi}}} \Bigg\{ & \int_{\boldsymbol{\epsilon}} \widetilde{\boldsymbol{\pi}}(\boldsymbol{\epsilon})  \sum_{i = 1}^{n} \Big( \lambda_{it}^+ \big(  \epsilon_{it}^+ + \Delta_i^+ V_t^{\boldsymbol{\pi}}(\boldsymbol{Q}_t, \boldsymbol{P}_t, S_t) \big) + \lambda_{it}^- \big( \epsilon_{it}^- + \Delta_i^- V_t^{\boldsymbol{\pi}}(\boldsymbol{Q}_t, \boldsymbol{P}_t, S_t) \Big) d\boldsymbol{\epsilon} \nonumber \\
        &- \gamma \int_{\boldsymbol{\epsilon}} \widetilde {\boldsymbol{\pi}}(\boldsymbol{\epsilon}_t) \log \widetilde{\boldsymbol{\pi}}(\boldsymbol{\epsilon}_t) d\boldsymbol{\epsilon}_t 
    \Bigg\}
    \end{align}
    then apply the $\widetilde{\boldsymbol{\pi}}$ at that time will result the following inequality,
    
    \begin{align}
    &  \int_{\boldsymbol{\epsilon}} \widetilde{\boldsymbol{\pi}}(\boldsymbol{\epsilon})  \sum_{i = 1}^{n} \Big( \lambda_{it}^+ \big(  \epsilon_{it}^+ + \Delta_i^+ V_t^{\boldsymbol{\pi}}(\boldsymbol{Q}_t, \boldsymbol{P}_t, S_t) \big) + \lambda_{it}^- \big( \epsilon_{it}^- + \Delta_i^- V_t^{\boldsymbol{\pi}}(\boldsymbol{Q}_t, \boldsymbol{P}_t, S_t) \Big) d\boldsymbol{\epsilon} \nonumber \\
    &- \gamma \int_{\boldsymbol{\epsilon}} \widetilde {\boldsymbol{\pi}}(\boldsymbol{\epsilon}_t) \log \widetilde{\boldsymbol{\pi}}(\boldsymbol{\epsilon}_t) d\boldsymbol{\epsilon}_t + \boldsymbol{Q}_t^{T} \big( \frac{\partial \boldsymbol{p_t}}{\partial t} + \frac{\partial^2 \boldsymbol p_t}{\partial S_t^2}\big) + \partial_t V_t^{\boldsymbol{\pi}}( \boldsymbol{Q}_t, \boldsymbol{P}_t, S_t) + \frac{\sigma^2 S_t^2}{2}\partial_{SS} V_t^{\boldsymbol{\pi}}( \boldsymbol{Q_t}, \boldsymbol{P}_t, S_t) \nonumber \\
    &+ (\nabla_{\boldsymbol{P}} V_t^{\boldsymbol{\pi}})^{T} \big( \frac{\partial \boldsymbol{p}_t}{\partial t}  + \frac{\sigma_t^2}{2} \frac{\partial^2 \boldsymbol{p}_t}{\partial S_t^2} \big) + \Big[\frac{\sigma_t^2 S_t^2}{2} (\frac{\partial \boldsymbol{p}_t}{\partial S_t})^T \boldsymbol{H}^{\boldsymbol{\pi}}_{\boldsymbol{P}}   (\frac{\partial \boldsymbol{p}_t}{\partial S_t}) + \frac{1}{2} \textbf{Tr} (\boldsymbol{V}_t^{T} \boldsymbol{H}^{\boldsymbol{\pi}}_{\boldsymbol{P}} \boldsymbol{V}_t) \Big] \geq 0
    \end{align}
    Thus, since we apply $\widetilde{\boldsymbol{\pi}}$ on $[t, s]$, which equivalently is optimizing the following quantity over every Infinitesimal in $[t, s]$
    \begin{align}
        \max_{\widetilde{\boldsymbol{\pi}}} \Bigg\{ & \int_{\boldsymbol{\epsilon}} \widetilde{\boldsymbol{\pi}}(\boldsymbol{\epsilon})  \sum_{i = 1}^{n} \Big( \lambda_{i}^+ \big(  \epsilon_{i}^+ + \Delta_i^+ V_u^{\boldsymbol{\pi}}(\boldsymbol{Q}, \boldsymbol{P}, S) \big) + \lambda_{i}^- \big( \epsilon_{i}^- + \Delta_i^- V_u^{\boldsymbol{\pi}}(\boldsymbol{Q}, \boldsymbol{P}, S) \Big) d\boldsymbol{\epsilon} \nonumber \\
        &- \gamma \int_{\boldsymbol{\epsilon}} \widetilde {\boldsymbol{\pi}}(\boldsymbol{\epsilon}) \log \widetilde{\boldsymbol{\pi}}(\boldsymbol{\epsilon}) d\boldsymbol{\epsilon} \nonumber 
    \Bigg\}
    \end{align}
    From the above argument, it is easy to derive the following inequality,  
    \begin{align}
        V_t^{\boldsymbol{\pi}}(\boldsymbol{Q}_t, \boldsymbol{P}_t, S_t) &\leq \int_t^s \int_{\boldsymbol{\epsilon}} \widetilde{\boldsymbol{\pi}}(\boldsymbol{\epsilon}_u) \sum_{i = 1}^{n} \big( \lambda_{iu}^+ \epsilon_{iu}^+ + \lambda_{iu}^- \epsilon_{iu}^- \big) d\boldsymbol{\epsilon}_u du - \gamma \int_t^s \int_{\boldsymbol{\epsilon}_u} \widetilde{\boldsymbol{\pi}}(\boldsymbol{\epsilon}_u) \log \widetilde{\boldsymbol{\pi}}(\boldsymbol{\epsilon}_u) d\boldsymbol{\epsilon}_u du \nonumber \\ 
        &+ \int_t^s  (\boldsymbol{Q}_u^{\boldsymbol{\widetilde{\boldsymbol{\pi}}}})^{T} \big( \frac{\partial \boldsymbol{p}_u}{\partial t} + \frac{\partial^2 \boldsymbol p_u}{\partial S^2}\big) du +  \mathbb E \Big[ V^{\boldsymbol{\pi}}_s(\boldsymbol{Q}^{\widetilde{\boldsymbol{\pi}}}_s, \boldsymbol{P}_s, S_s) \ \Big | \ t, \boldsymbol{Q}_t, \boldsymbol{P}_t, S_t \Big]
    \end{align}
    Let $s = T$, then $V_T^{\boldsymbol{\pi}}(\boldsymbol{Q}_T^{\widetilde{\boldsymbol{\pi}}},\boldsymbol{P}_T, S_T) = V_T^{\widetilde{\boldsymbol{\pi}}}(\boldsymbol{Q}_T^{\widetilde{\boldsymbol{\pi}}}, \boldsymbol{P}_T, S_T) = -\psi_0 \sum_{i = 1}^{n} (P_{iT} Q^{\widetilde{\boldsymbol{\pi}}}_{iT})^2$, then 
    \begin{align}
         V_t^{\boldsymbol{\pi}}(\boldsymbol{Q}_t, \boldsymbol{P}_t, S_t) &\leq \int_t^T \int_{\boldsymbol{\epsilon}} \widetilde{\boldsymbol{\pi}}(\boldsymbol{\epsilon}_u) \sum_{i = 1}^{n} \big( \lambda_{iu}^+ \epsilon_{iu}^+ + \lambda_{iu}^- \epsilon_{iu}^- \big) d\boldsymbol{\epsilon}_u du - \gamma \int_t^T \int_{\boldsymbol{\epsilon}_u} \widetilde{\boldsymbol{\pi}}(\boldsymbol{\epsilon}_u) \log \widetilde{\boldsymbol{\pi}}(\boldsymbol{\epsilon}_u) d\boldsymbol{\epsilon}_u du \nonumber \\ 
        &+ \int_t^T (\boldsymbol{Q}_u^{\boldsymbol{\widetilde{\boldsymbol{\pi}}}})^{T} \big( \frac{\partial \boldsymbol{p}_u}{\partial t} + \frac{\partial^2 \boldsymbol p_u}{\partial S^2}\big) du +  \mathbb E \Big[ V^{\boldsymbol{\pi}}_T(\boldsymbol{Q}^{\widetilde{\boldsymbol{\pi}}}_T, \boldsymbol{P}_T, S_T) \ \Big | \ t, \boldsymbol{Q}_t, \boldsymbol{P}_t, S_t \Big] \nonumber \\
        & = V_t^{\widetilde{\boldsymbol{\pi}}}(\boldsymbol{Q}_t, \boldsymbol{P}_t, S_t)
    \end{align}
    
\section{Proof of Optimal Policy Convergence}\label{Optimal Convergence}
Consider the policy $\boldsymbol{\pi}_{\infty}$ derived from $V^{\infty}$, which is 
    \begin{align}
        \boldsymbol{\pi}_{\infty} \sim \frac{\exp \Big \{ \frac{1}{\gamma} \sum_{i = 1}^{n} \lambda_{it}^+ \big(  \epsilon_{it}^+ + \Delta_i^+ V_t^{\infty}(\boldsymbol{Q}_t, \boldsymbol{P}_t, S_t) \big) + \sum_{i = 1}^{n} \lambda_{it}^- \big( \epsilon_{it}^- + \Delta_i^- V_t^{\infty}(\boldsymbol{Q}_t, \boldsymbol{P}_t, S_t)\big)    \Big\}}{\int_{\boldsymbol{\epsilon_t}}\exp \Big \{ \frac{1}{\gamma} \sum_{i = 1}^{n} \lambda_{it}^+ \big(  \epsilon_{it}^+ + \Delta_i^+ V_t^{\infty}(\boldsymbol{Q}_t, \boldsymbol{P}_t, S_t) \big) + \sum_{i = 1}^{n} \lambda_{it}^- \big( \epsilon_{it}^- + \Delta_i^- V_t^{\infty}(\boldsymbol{Q}_t, \boldsymbol{P}_t, S_t)\big)    \Big\} d\boldsymbol{\epsilon_t}}
    \end{align}
    For any other admissible policy $\boldsymbol{\pi}$, since $\boldsymbol{\pi}_{\infty}$ is the maximizer of the following optimization, 
        \begin{align}
        \max_{\boldsymbol{\pi}} \Bigg\{ &\int_{\boldsymbol{\epsilon}} {\boldsymbol{\pi}}(\boldsymbol{\epsilon})  \sum_{i = 1}^{n} \Big( \lambda_{it}^+ \big(  \epsilon_{it}^+ + \Delta_i^+ V_t^{\infty}(\boldsymbol{Q}_t, \boldsymbol{P}_t, S_t) \big) + \lambda_{it}^- \big( \epsilon_{it}^- + \Delta_i^- V_t^{\infty}(\boldsymbol{Q}_t, \boldsymbol{P}_t, S_t) \Big) d\boldsymbol{\epsilon} \nonumber \\ 
        &- \gamma \int_{\boldsymbol{\epsilon}} \widetilde {{\pi}}(\boldsymbol{\epsilon}_t) \log {\boldsymbol{\pi}}(\boldsymbol{\epsilon}_t) d\boldsymbol{\epsilon}_t \nonumber 
    \Bigg\}
    \end{align}
    In addition, the following equation holds, 
    \begin{align}
         & \int_{\boldsymbol{\epsilon}}{\boldsymbol{\pi}_{\infty}}(\boldsymbol{\epsilon})  \sum_{i = 1}^{n} \Big( \lambda_{it}^+ \big(  \epsilon_{it}^+ + \Delta_i^+ V_t^{\infty}(\boldsymbol{Q}_t, \boldsymbol{P}_t, S_t) \big) + \lambda_{it}^- \big( \epsilon_{it}^- + \Delta_i^- V_t^{\infty}(\boldsymbol{Q}_t, \boldsymbol{P}_t, S_t) \Big) d\boldsymbol{\epsilon}\nonumber \\
         &- \gamma \int_{\boldsymbol{\epsilon}} {\boldsymbol{\pi}_{\infty}}(\boldsymbol{\epsilon}) \log {\boldsymbol{\pi}_{\infty}}(\boldsymbol{\epsilon}_t) d\boldsymbol{\epsilon}_t + \boldsymbol{Q}_t^{T} \big( \frac{\partial \boldsymbol{p_t}}{\partial t} + \frac{\partial^2 \boldsymbol p_t}{\partial S_t^2}\big) + \partial_t V_t^{\infty}( \boldsymbol{Q}_t, \boldsymbol{P}_t, S_t) + \frac{\sigma^2 S_t^2}{2}\partial_{SS} V_t^{\infty}( \boldsymbol{Q_t}, \boldsymbol{P}_t, S_t) \nonumber \\
    &+ (\nabla_{\boldsymbol{P}} V_t^{\infty})^{T} \big( \frac{\partial \boldsymbol{p}_t}{\partial t}  + \frac{\sigma_t^2}{2} \frac{\partial^2 \boldsymbol{p}_t}{\partial S_t^2} \big) + \Big[\frac{\sigma_t^2 S_t^2}{2} (\frac{\partial \boldsymbol{p}_t}{\partial S_t})^T \boldsymbol{H}^{\infty}_{\boldsymbol{P}}   (\frac{\partial \boldsymbol{p}_t}{\partial S_t}) + \frac{1}{2} \textbf{Tr} (\boldsymbol{V}_t^{T} \boldsymbol{H}^{\infty}_{\boldsymbol{P}} \boldsymbol{V}_t) \Big] = 0
    \end{align}
    Then for arbitrary policy $\boldsymbol{\pi}$, there is an inequality
    \begin{align}
        & \int_{\boldsymbol{\epsilon}}{\boldsymbol{\pi}}(\boldsymbol{\epsilon})  \sum_{i = 1}^{n} \Big( \lambda_{it}^+ \big(  \epsilon_{it}^+ + \Delta_i^+ V_t^{\infty}(\boldsymbol{Q}_t, \boldsymbol{P}_t, S_t) \big) + \lambda_{it}^- \big( \epsilon_{it}^- + \Delta_i^- V_t^{\infty}(\boldsymbol{Q}_t, \boldsymbol{P}_t, S_t) \Big) d\boldsymbol{\epsilon} \nonumber \\
        &- \gamma \int_{\boldsymbol{\epsilon}} {\boldsymbol{\pi}}(\boldsymbol{\epsilon}) \log {\boldsymbol{\pi}}(\boldsymbol{\epsilon}_t) d\boldsymbol{\epsilon}_t + \boldsymbol{Q}_t^{T} \big( \frac{\partial \boldsymbol{p_t}}{\partial t} + \frac{\partial^2 \boldsymbol p_t}{\partial S_t^2}\big) + \partial_t V_t^{\infty}( \boldsymbol{Q}_t, \boldsymbol{P}_t, S_t) + \frac{\sigma^2 S_t^2}{2}\partial_{SS} V_t^{\infty}( \boldsymbol{Q_t}, \boldsymbol{P}_t, S_t) \nonumber \\
    &+ (\nabla_{\boldsymbol{P}} V_t^{\infty})^{T} \big( \frac{\partial \boldsymbol{p}_t}{\partial t}  + \frac{\sigma_t^2}{2} \frac{\partial^2 \boldsymbol{p}_t}{\partial S_t^2} \big) + \Big[\frac{\sigma_t^2 S_t^2}{2} (\frac{\partial \boldsymbol{p}_t}{\partial S_t})^T \boldsymbol{H}^{\infty}_{\boldsymbol{P}}   (\frac{\partial \boldsymbol{p}_t}{\partial S_t}) + \frac{1}{2} \textbf{Tr} (\boldsymbol{V}_t^{T} \boldsymbol{H}^{\infty}_{\boldsymbol{P}} \boldsymbol{V}_t) \Big] \leq 0
    \end{align}
    Following the similar argument in theorem 1, there is inequality,
    \begin{align}
        V^{\infty}_t (\boldsymbol{Q}_t, \boldsymbol{P}_t, S_t) \geq V^{\boldsymbol{\pi}}_t (\boldsymbol{Q}_t, \boldsymbol{P}_t, S_t)
    \end{align}

\end{document}